 \documentclass[10pt,pra,aps,a4paper,oneside,twocolumn,superscriptaddress,showpacs]{revtex4-1}

\usepackage[english]{babel}
\usepackage[latin1]{inputenc}
\usepackage[T1]{fontenc}
\usepackage{lmodern}
\usepackage{ulem}
\usepackage{dcolumn}
\usepackage{subfigure}
\usepackage{footnote}
\usepackage{multirow}
\usepackage{amsmath,amsfonts,amssymb}

\usepackage{hyperref}
\usepackage{cleveref}

\usepackage{graphicx}  % standard LaTeX graphics tool
\usepackage{latexsym}   % useful for coding complex math
\usepackage{color}

\begin{document}

\title{Electromagnetic energy and negative asymmetry parameter in coated magneto-optical cylinders: Applications to tunable light transport in disordered systems}

\author{\firstname{Tiago}  J. \surname{Arruda}}
\email{tiagoarruda@pg.ffclrp.usp.br}
\affiliation{Faculdade de Filosofia,~Ci\^encias e Letras de Ribeir\~ao
Preto, Universidade de S\~ao Paulo, 14040-901 Ribeir\~ao
Preto, S\~ao Paulo, Brazil}

\author{\firstname{Alexandre} S. \surname{Martinez}}
%\email{asmartinez@ffclrp.usp.br}
\affiliation{Faculdade de Filosofia,~Ci\^encias e Letras de Ribeir\~ao
Preto, Universidade de S\~ao Paulo, 14040-901 Ribeir\~ao
Preto, S\~ao Paulo, Brazil}
\affiliation{National Institute of Science and Technology in
Complex Systems, 22290-180 Rio de Janeiro, Rio de Janeiro, Brazil}

\author{\firstname{Felipe}  A. \surname{Pinheiro}}
\affiliation{Instituto de F\'{i}sica, Universidade Federal do Rio de Janeiro, 21941-972 Rio de Janeiro, Rio de Janeiro, Brazil}
%\affiliation{Optoelectronics Research Centre and Centre for Photonic Metamaterials, University of Southampton, Highfield, Southampton SO17 1BJ, United Kingdom}

\begin{abstract}

We investigate electromagnetic scattering of normally irradiated gyrotropic, magneto-optical core-shell cylinders using Lorenz-Mie theory. 
A general expression for time-averaged electromagnetic energy inside a coated gyroelectric and gyromagnetic  scatterer is derived. 
Using realistic material parameters for a silica core and InSb shell, we calculate the stored electromagnetic energy and the scattering anisotropy. 
We show that the application of an external magnetic field along the cylinder axis induces a drastic decrease in electromagnetic absorption in a frequency range in the terahertz, where absorption is maximal in the absence of the magnetic field.
We demonstrate not only that the scattering anisotropy can be externally tuned by applying a magnetic field, but also that it reaches negative values in the terahertz range even in the dipolar regime. 
We also show that this preferential backscattering response results in an anomalous regime of multiple light scattering from a collection of magneto-optical core-shell cylinders, in which the  extinction mean free path is longer than the transport mean free path. 
By additionally calculating the energy-transport velocity and diffusion coefficient, we demonstrate an unprecedented degree of external control of multiple light scattering, which can be achieved by either applying an external magnetic field or varying the temperature.   

\end{abstract}


\maketitle

%\tableofcontents

\section{Introduction}

Electromagnetic (EM) scattering by small particles with sizes of the order of the wavelength has many applications not only in different areas of physics, such as meteorology, optical communications, sensing, and astrophysics~\cite{Bohren}, but also in interdisciplinary fields, such as biophysics~\cite{Huschka}. 
The advent of plasmonics and metamaterials has allowed for the discovery of novel scattering phenomena, which do not exist in naturally occurring materials such as plasmonic cloaking~\cite{Alu2008,Edwards,Wilton-prl}, unconventional Fano resonances~\cite{Tribelsky,Tiago_pra,Chen_gao,Jordi}, artificial magnetism~\cite{Miromagnetic,Kuznetsov}, superscattering~\cite{Ruan}, and the unprecedented control of the scattering directionality~\cite{Liu,Liu2013}. 

In particular, controlling the scattering direction crucially depends on the design optimization of the electric and magnetic responses of small particles. 
Most approaches to controlling the scattering direction rely on tailoring the electric structures of nanoparticles, as typically the electric response is dominant in natural media at optical frequencies~\cite{Bohren,Boris}. 
However, efforts have been made to propose and design metamaterials that support both electric and magnetic dipolar resonances, such as spheres~\cite{Liu,Etxarri} and high permittivity cylinders~\cite{Liu2013}. 
These efforts have allowed the achievement of zero-backward scattering and near-zero-forward scattering conditions (known as Kerker conditions), first theoretically predicted for hypothetical particles exhibiting both electric and magnetic dipolar resonances~\cite{Kerker}, and highly directional EM scattering~\cite{Staude,Vesperinas,Coenen,Fu,Zambrana,Hancu}. 
The observation of Kerker conditions relies on the interference of electric and magnetic dipoles in nanostructures~\cite{Person,Geffrin}. 
Alternative theoretical approaches for Kerker conditions, involving electric dipoles and quadrupoles, also exist~\cite{Alaee}. 
Recently, both broadband zero backward and near-zero forward scattering have been obtained even beyond the dipole limit, relaxing design constraints in practical experiments~\cite{Li}. 
Other mechanisms for achieving directional light scattering have also been proposed using magneto-optical materials~\cite{Wilton_josa,Hall}.

In addition, achieving preferential backward scattering remains a challenge and has many applications in multiple light scattering. 
Preferential backscattering, which hardly occurs in natural media, is characterized by negative values of the asymmetry parameter, the average of the cosine of the scattering angle $\langle \cos \theta \rangle$. 
Indeed, for small particles in the Rayleigh regime, scattering is dipolar so that $\langle \cos \theta \rangle\approx 0$. 
In contrast, Mie particles scatter strongly in the forward direction, $\langle \cos \theta \rangle \approx 1$. 
Negative $\langle \cos \theta \rangle$ has been reported in ferromagnetic particles~\cite{Felipe_prl} and lossless dielectric nanospheres made of moderate permittivity materials, such as silicon or germanium nanospheres in the infrared region~\cite{Medina2012}. 
In these cases, negative asymmetry parameters have been show to lead to an unusual regime in multiple light scattering, in which the scattering mean free path is larger than the transport mean free path~\cite{Medina2012}. 
This peculiar transport regime has also been demonstrated for multiple scattering systems with correlated disorder under certain conditions~\cite{Conley}. 

In this paper we propose an alternative, versatile strategy not only to achieve preferential backscattering (negative $\langle \cos \theta \rangle$) but also to control the scattering direction with an external parameter, which can be either an applied magnetic field or the temperature. 
We demonstrate that these effects can be obtained by investigating light scattering in dielectric cylinders coated with a gyroelectric, magneto-optical shell. 
For concreteness, we consider realistic material parameters: a silica (SiO$_2$) core and an indium antimonide (InSb) shell, which has a strong magneto-optical and temperature dependence of the dielectric function on the response in the far-infrared~ \cite{Madelung,Zimpel,Howells}. 
Specifically, in the framework of the Lorenz-Mie theory we show that the asymmetry scattering factor for this system can be tuned from negative to positive values by applying moderate magnetic fields ${\mathbf B}$ at the Voigt configuration. 
We derive a closed analytical expression for the time-averaged EM energy stored inside the core-shell cylinder for both TE and TM polarizations to show that the application of an external magnetic field induces a gap in EM absorption in the terahertz (THz) regime. 
We also consider multiple light scattering by an assembly of coated-cylinders and calculate the transport mean-free path, energy-transport velocity and light diffusion coefficient. 
This analysis reveals that the unusual light transport regime where the scattering mean free path is longer than the transport mean free path can be reached by varying either the temperature or the external magnetic field.   

This paper is organized as follows.
In Sec.~\ref{theory}, we present the system and fundamental quantities regarding the Lorenz-Mie theory in a consistent notation.
Our expressions are general and can be applied to arbitrary core-shell cylinders containing gyroelectric and gyromagnetic materials normally irradiated with plane waves in TE or TM polarizations.
 The main analytical result is presented in Sec.~{\ref{results}, in which we calculate the time-averaged internal energy and its relation to the absorption cross section, generalizing previous results obtained for isotropic cylinders.
In Sec.~\ref{graphics}, we study both single scattering and diffusion of EM waves in two-dimensional (2D) disordered media consisting of (SiO$_2$) core-shell (InSb) cylinders and discuss some applications.
Finally, in Sec.~\ref{conclusion}, we summarize our results and conclude.

\section{The Lorenz-Mie theory for gyrotropic coated cylinders}
\label{theory}

To  investigate magneto-optical effects in 2D media, we consider a center-symmetric, infinitely long, gyrotropic core-shell cylinder, with inner radius $a$ and outer radius $b$, normally irradiated with monochromatic plane waves.
This scatterer is embedded in an isotropic lossless medium $(\varepsilon_0,\mu_0)$, where $\varepsilon$ and $\mu$ are the dielectric permittivity and magnetic permeability, respectively.
The material parameters of the coated cylinder are $(\overleftrightarrow{\boldsymbol{\varepsilon}_1},\overleftrightarrow{\boldsymbol{\mu}_1})$ for the core ($0< r\leq a$) and $(\overleftrightarrow{\boldsymbol{\varepsilon}_2},\overleftrightarrow{\boldsymbol{\mu}_2})$ for the shell ($a<r\leq b$), where the gyroelectric and gyromagnetic tensors are, respectively,
\begin{align}
    \overleftrightarrow{\boldsymbol{\varepsilon}_q}&= \begin{pmatrix} 
		                                                                             \varepsilon_{xx} & \varepsilon_{xy} & 0 \\
                                                                                 \varepsilon_{yx}&  \varepsilon_{yy} & 0\\
                                                                                  0 & 0 &\varepsilon_{zz}\end{pmatrix}=
																																									\begin{pmatrix} 
																																										\varepsilon_q^{\perp} & \imath\gamma_q & 0 \\
                                                                                    -\imath\gamma_q&   \varepsilon_q^{\perp}      & 0\\
                                                                                    0 & 0 &\varepsilon_q^{||}\end{pmatrix}
 ,\label{epsilon}\\
    \overleftrightarrow{\boldsymbol{\mu}_q}&= \begin{pmatrix} 
		                                                                             \mu_{xx} & \mu_{xy} & 0 \\
                                                                                 \mu_{yx}&  \mu_{yy} & 0\\
                                                                                  0 & 0 &\mu_{zz}\end{pmatrix}=
																																									\begin{pmatrix} 
																																										\mu_q^{\perp} & \imath\eta_q & 0 \\
                                                                                    -\imath\eta_q&   \mu_q^{\perp}      & 0\\
                                                                                    0 & 0 &\mu_q^{||}\end{pmatrix}\ .\label{mu}
\end{align}

Considering the time harmonic dependence $e^{-\imath\omega t}$, where $\omega$ is the angular frequency, one has the curl Maxwell equations $\nabla\times (\mathbf{E},\mathbf{H})= \imath\omega (\overleftrightarrow{\boldsymbol{\mu}}\cdot\mathbf{H},-\overleftrightarrow{\boldsymbol{\varepsilon}}\cdot\mathbf{E})$.
In cylindrical coordinates $(r,\phi,z)$, there are two irradiation schemes with analytical solutions, as depicted in Fig.~\ref{fig1}: the TM polarization or $p$ wave ($\mathbf{H}||\hat{\mathbf{z}}$), which provides the field components $E_r$ and $E_{\phi}$ in terms of partial derivatives of $H_z$; and the TE polarization or $s$ wave ($\mathbf{E}||\hat{\mathbf{z}}$), which leads to $H_r$ and $H_{\phi}$ in terms of partial derivatives of $E_z$. 

\begin{figure}[htbp]
\centerline{\includegraphics[width=\columnwidth]{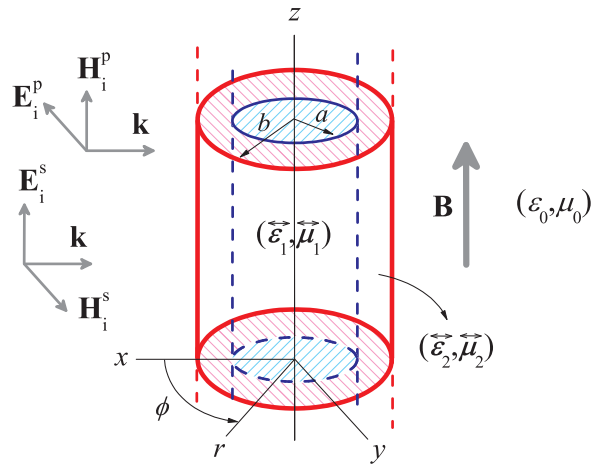}}
\caption{A center-symmetric core-shell, infinitely long circular cylinder normally irradiated with plane waves in the Voigt configuration $(\mathbf{B}\perp\mathbf{k})$.
The core has optical properties $({\boldsymbol{\varepsilon}_1},{\boldsymbol{\mu}_1})$ and radius $a$, whereas the shell has $({\boldsymbol{\varepsilon}_2},{\boldsymbol{\mu}_2})$ and radius $b$.
The surrounding medium is the vacuum $(\varepsilon_0,\mu_0)$. 
The incident EM fields $(\mathbf{E}_{\rm i}^{\rm p},\mathbf{H}_{\rm i}^{\rm p})$ and $(\mathbf{E}_{\rm i}^{\rm s},\mathbf{H}_{\rm i}^{\rm s})$ are on $p$ or $s$ polarizations, respectively. 
The applied external magnetic field satisfies $|\mathbf{B}|\gg|\mu_0\mathbf{H}_{\rm i}|$.}\label{fig1}
\end{figure}

For both polarizations, we define the following quantities, used throughout this text:
\begin{align}
{\rm TM\  (p)}&\Rightarrow 
\begin{cases}
\varepsilon_q^{\rm p}\equiv \varepsilon_q^{\perp}[1-(\beta_q^{\rm p})^2]\ ,\\
\mu_q^{\rm p}\equiv \mu_q^{||}\ ,\\
\beta_q^{\rm p}\equiv \gamma_q/\varepsilon_q^{\perp}\ ;
\end{cases}\label{tm-p}\\
{\rm TE\  (s)}&\Rightarrow 
\begin{cases}
\varepsilon_q^{\rm s}\equiv \varepsilon_q^{||}\ ,\\
\mu_q^{\rm s}\equiv \mu_q^{\perp} [1-(\beta_q^{\rm s})^2]\ ,\\
\beta_q^{\rm s}\equiv \eta_q/\mu_q^{\perp}\ .
\end{cases}\label{te-s}
\end{align}
With this simplified notation, for both $p$ and $s$ waves,  $\beta_q=0$ (also known as the Voigt parameter) describes an isotropic 2D medium with $(\varepsilon_q,\mu_q)$.
In the following, we focus on $p$-polarized waves and cylinders composed of gyroelectric materials, Eq.~(\ref{epsilon}).
The discussion for $s$ polarization is  analogous.

\subsection{Electric and magnetic fields for $p$ waves}

The EM wave impinging on the cylinder is set as a monochromatic wave propagating with wave vector $\mathbf{k}=-k\hat{\mathbf{x}}$ and time harmonic dependence $e^{-\imath \omega t}$. 
The scatterer geometry is depicted in Fig.~\ref{fig1}.
For $p$-polarized waves in cylindrical coordinate system $(r,\phi,z)$, we have the \textit{ansatz} $[\mathbf{E}_{\rm i}^{\rm p}(r,\phi),\mathbf{H}_{\rm i}^{\rm p}(r,\phi)]=(-E_0\hat{\mathbf{y}},H_0\hat{\mathbf{z}})e^{-\imath kr \cos\phi}$, where the electric and magnetic amplitudes are related by $E_0=H_0\sqrt{\varepsilon_0/\mu_0}$.

Expanding the incident EM field  in vector cylindrical harmonics, we obtain for $r> b$ the nonvanishing components
\begin{align}
    E_{{\rm i}r}^{\rm p}&=-\sum_{n=-\infty}^{\infty}E_n n\frac{J_n(kr)}{kr}e^{\imath n\phi}\ ,\label{Eir}\\
    E_{{\rm i}\phi}^{\rm p}&=-\imath\sum_{n=-\infty}^{\infty}E_n J_n'(kr)e^{\imath n\phi}\ ,\label{Eiphi}\\
    H_{{\rm i}z}^{\rm p}&=\frac{k}{\omega\mu_0}\sum_{n=-\infty}^{\infty}E_n J_n(kr)e^{\imath n\phi}\ ,\label{Hiz}
\end{align}
where $E_n=E_0(-\imath)^n $, $k^2=\omega^2\epsilon_0\mu_0$, and $J_n$ is the cylindrical Bessel function.
As a consequence, the nonvanishing components of the EM field scattered by the cylinder are, for $r>b$~\cite{Bohren}, 
\begin{align}
    E_{{\rm s}r}^{\rm p}&=-\sum_{n=-\infty}^{\infty}E_n a_n^{\rm p}n\frac{H_n^{(1)}(kr)}{kr}e^{\imath n\phi}\ ,\label{Esr}\\
    E_{{\rm s}\phi}^{\rm p}&=-\imath\sum_{n=-\infty}^{\infty}E_n a_n^{\rm p}H_n'^{(1)}(kr)e^{\imath n\phi}\ ,\label{Esphi}\\
    H_{{\rm s}z}^{\rm p}&=\frac{k}{\omega\mu_0}\sum_{n=-\infty}^{\infty}E_n a_n^{\rm p}H_n^{(1)}(kr)e^{\imath n\phi}\ ,\label{Hsz}
\end{align}
where $a_n$ is the scattering coefficient and $H_n^{(1)}$ is the cylindrical Hankel function of the first kind.
The form of the scattering coefficient depends on the material properties of the scatterer.

From Maxwell's equations, one can show that the magnetic field $\mathbf{H}_q=H_{qz}\hat{\mathbf{z}}$  within the scatterer must satisfy the following Helmholtz equation~\cite{Monzon,Wilton_josa}: $(\mathbf{\nabla}^2 + k_q^{\rm p})H_{qz}=0$, where $(k_q^{\rm p})^2=\omega^2\epsilon_q^{\rm p}\mu_q^{\rm p}$.
The remaining EM field components are calculated from $(k_q^{\rm p})^2 E_{q r} = \imath\omega\mu_q^{\rm p} [\imath \beta_q^{\rm p}\partial/\partial {r}+ (1/r)\partial/\partial {\phi}] H_{qz}$ and $(k_q^{\rm p})^2 E_{q \phi} = -\imath\omega\mu_q^{\rm p} [\partial/\partial {r} - (\imath\beta_q^{\rm p}/r)\partial/\partial{\phi}] H_{qz}$.
Explicitly, we have for the core region, $q=1$ $(0<r\leq a)$,
\begin{align}
    E_{1r}^{\rm p}&=-\sum_{n=-\infty}^{\infty}E_n b_n^{\rm p}\widetilde{\mathcal{J}}_n(k_1^{\rm p}r,\beta_1^{\rm p})e^{\imath n\phi}\ ,\label{E1r}\\
    E_{1\phi}^{\rm p}&=-\imath\sum_{n=-\infty}^{\infty}E_n b_n^{\rm p}{\mathcal{J}}_n(k_1^{\rm p}r,\beta_1^{\rm p})e^{\imath n\phi}\ ,\label{E1phi}\\
    H_{1z}^{\rm p}&=\frac{k_1^{\rm p}}{\omega\mu_1^{\rm p}}\sum_{n=-\infty}^{\infty} E_n b_n^{\rm p}J_n(k_1^{\rm p}r)e^{\imath n\phi}\ ,\label{H1z}
		\end{align}
		where, for the sake of simplicity, we define $\mathcal{J}_n(\rho,\beta)\equiv J_n'(\rho)+\beta n{J_n(\rho)}/{\rho}$ and $\widetilde{\mathcal{J}}_n(\rho,\beta)\equiv \beta J_n'(\rho)+ n{J_n(\rho)}/{\rho}$;
and, for the shell region, $q=2$ $(a\leq r\leq b)$,	
\begin{align}
		    E_{2r}^{\rm p}=&-\sum_{n=-\infty}^{\infty}E_n \left[c_n^{\rm p}\widetilde{\mathcal{J}}_n(k_2^{\rm p}r,\beta_2^{\rm p})+d_n^{\rm p}\widetilde{\mathcal{Y}}_n(k_2^{\rm p}r,\beta_2^{\rm p})\right]e^{\imath n\phi}\ ,\label{E2r}\\
    E_{2\phi}^{\rm p}=&-\imath\sum_{n=-\infty}^{\infty}E_n \left[c_n^{\rm p}{\mathcal{J}}_n(k_2^{\rm p}r,\beta_2^{\rm p})+d_n^{\rm p}{\mathcal{Y}}_n(k_2^{\rm p}r,\beta_2^{\rm p})\right]e^{\imath n\phi}\ ,\label{E2phi}\\
    H_{2z}^{\rm p}=&\frac{k_2^{\rm p}}{\omega\mu_2^{\rm p}}\sum_{n=-\infty}^{\infty} E_n \left[c_n^{\rm p}J_n(k_2^{\rm p}r) + d_n^{\rm p}Y_n(k_2^{\rm p}r)\right]e^{\imath n\phi}\ ,\label{H2z}
\end{align}
where $\mathcal{Y}_n(\rho,\beta)\equiv Y_n'(\rho)+\beta n{Y_n(\rho)}/{\rho}$ and $\widetilde{\mathcal{Y}}_n(\rho,\beta)\equiv \beta Y_n'(\rho)+ n{Y_n(\rho)}/{\rho}$, with $Y_n$ being the cylindrical Neumann function. 
			
The Lorenz-Mie coefficients $a_n^{\rm p}$, $b_n^{\rm p}$, $c_n^{\rm p}$ and $d_n^{\rm p}$ are obtained by imposing the boundary conditions at $r=a$ and $r=b$, reading:
\begin{align}
   a_n^{\rm p} &= \frac{\widetilde{m}_2^{\rm p}J_n'(y)\left[J_n(m_2^{\rm p}y) - \mathcal{A}_n^{\rm p} Y_n(m_2^{\rm p}y)\right] - J_n(y){\alpha}_n^{\rm p} }
           {\widetilde{m}_2^{\rm p}H_n'^{(1)}(y)\left[J_n(m_2^{\rm p}y) - \mathcal{A}_n^{\rm p} Y_n(m_2^{\rm p}y)\right]- H_n^{(1)}(y){\alpha}_n^{\rm p}}\ ,\label{an}\\
			b_n^{\rm p} &= \frac{\widetilde{m}_2^{\rm p} c_n^{\rm p} \left[ J_n(m_2^{\rm p}x) - \mathcal{A}_n^{\rm p} Y_n(m_2^{\rm p}x)\right]}{\widetilde{m}_1^{\rm p} J_n(m_1^{\rm p}x)}\ ,\label{dn}\\
    c_n^{\rm p}&= \frac{2\imath/(\pi y)}{\widetilde{m}_2^{\rm p}H_n'^{(1)}(y)\left[J_n(m_2^{\rm p}y) - \mathcal{A}_n^{\rm p} Y_n(m_2^{\rm p}y)\right] - H_n^{(1)}(y){\alpha}_n^{\rm p}}\ ,\label{gn}\\
		d_n^{\rm p} & = -\mathcal{A}_n^{\rm p} c_n^{\rm p}\ ,\label{wn}
	\end{align}
	where the auxiliary functions are 
	\begin{align*}
	\alpha_n^{\rm p} &= \mathcal{J}_n(m_2^{\rm p}y,\beta_2^{\rm p}) - \mathcal{A}_n^{\rm p} \mathcal{Y}_n(m_2^{\rm p}y,\beta_2^{\rm p})\ ,\\
		\mathcal{A}_n^{\rm p} &= \frac{\widetilde{m}_1^{\rm p} J_n(m_1^{\rm p}x)\mathcal{J}_n(m_2^{\rm p}x,\beta_2^{\rm p})-\widetilde{m}_2^{\rm p}\mathcal{J}_n(m_1^{\rm p}x,\beta_1^{\rm p})J_n(m_2^{\rm p}x)}{\widetilde{m}_1^{\rm p}J_n(m_1^{\rm p}x)\mathcal{Y}_n(m_2^{\rm p}x,\beta_2^{\rm p})-\widetilde{m}_2^{\rm p}\mathcal{J}_n(m_1^{\rm p}x,\beta_1^{\rm p})Y_n(m_2^{\rm p}x)}\ ,
		\end{align*}
		with size parameters $x=ka$ and $y=kb$.
		The relative refractive and impedance indices are $m_q^{\rm p}=k_q^{\rm p}/k=\sqrt{\varepsilon_q^{\rm p}\mu_q^{\rm p}/(\varepsilon_0\mu_0)}$ and $\widetilde{m}_q^{\rm p}=\sqrt{\varepsilon_q^{\rm p}\mu_0/(\varepsilon_0\mu_q^{\rm p})}$, respectively ($m_q=\widetilde{m}_q$ if $\mu_q=\mu_0$~\cite{Felipe_prl}).		
Notice that parity symmetry $a_{-n}=a_n$, $b_{-n}=b_n$, $c_{-n}=c_n$ and $d_{-n}=d_n$ only holds if $\beta_1=\beta_2=0$, which retrieves the isotropic result for the TM mode~\cite{Tiago_JOSA2010,Tiago_joa,Tiago_JOSA2013}.

The corresponding expressions for $s$-polarized waves are analogous to the ones above.
For the sake of completeness, the multipole expansions and the Lorenz-Mie coefficients $(a_n^{\rm s},b_n^{\rm s},c_n^{\rm s},d_n^{\rm s})$ are presented in Appendix~\ref{TEmode}.
These expressions are necessary to study cylinders composed of gyromagnetic materials, Eq.~(\ref{mu}).

\subsection{Lorenz-Mie efficiencies and multiple scattering}

The extinction and scattering efficiencies  for cylindrical scatterers at normal incidence are directly calculated via $Q_{\rm sca}=(2/y)\sum_{n=-\infty}^{\infty}|a_n|^2$ and $Q_{\rm ext}=(2/y)\sum_{n=-\infty}^{\infty}{\rm Re}(a_n)$, respectively,  where $y=kb$ is the size parameter of the outer cylinder.
They are defined as the respective cross section of a segment $L\gg b$ of the infinite cylinder in units of the geometrical cross section $2bL$.
Rewriting these efficiencies to consider sums for $n\geq 1$, one has 
\begin{align}
    Q_{\rm sca} &= \frac{2}{y}\left[\left|a_0\right|^2+\sum_{n=1}^{\infty}\left(\left|a_{-n}\right|^2+\left|a_n\right|^2\right)\right]\ ,\label{Qsca}\\
    Q_{\rm ext} &= \frac{2}{y}{\rm Re}\left[a_0+\sum_{n=1}^{\infty}\left(a_{-n} + a_n\right)\right]\ ,\label{Qext}
\end{align}
with $Q_{\rm abs} = Q_{\rm ext}-Q_{\rm sca}$ being the absorption efficiency. 
The differential scattering efficiency reads 
\begin{align}
    \frac{\partial Q(\phi)}{\partial\phi}  =\frac{2}{\pi y}\left|a_0 + \sum_{n=1}^{\infty}\left(a_{-n} e^{\imath n\theta} + a_n e^{-\imath n\theta}\right)\right|^2\ ,\label{Qdif}
 \end{align}
where $\theta = \pi-\phi$ is the scattering angle, so that $\theta=0^{\rm o}$ corresponds to forward scattering and $\theta=180^{\rm o}$ corresponds to backscattering.
Here we consider the same convention as in Refs.~\cite{Kivshar_cross} and \cite{Wilton_josa}, so that one obtains $Q_{\rm sca}$ by integrating Eq.~(\ref{Qdif}) in the range $[0,\pi]$ instead of $[0,2\pi]$~\cite{Hulst}.
The efficiencies for $p$  and $s$ polarizations are obtained by considering $a_{n}^{\rm  p}$ and $a_{n}^{\rm s}$, respectively, where one must define the quantities $(\varepsilon_q,\mu_q,\beta_q)$ according to relations~(\ref{tm-p}) or (\ref{te-s}).

Some quantities calculated in the single scattering approach can be used to study multiple-scattering properties in the diffusive regime and for low concentrations of scatterers~\cite{Tiggelen1,Tiggelen2}.
In this regime, the scattering mean free path $\ell_{\rm sca}$  is comparable to the size of the system and suffices $k\ell_{\rm sca}\gg1$.
This situation is depicted in Fig.~\ref{fig2}.

\begin{figure}[htbp]
\centerline{\includegraphics[width=\columnwidth]{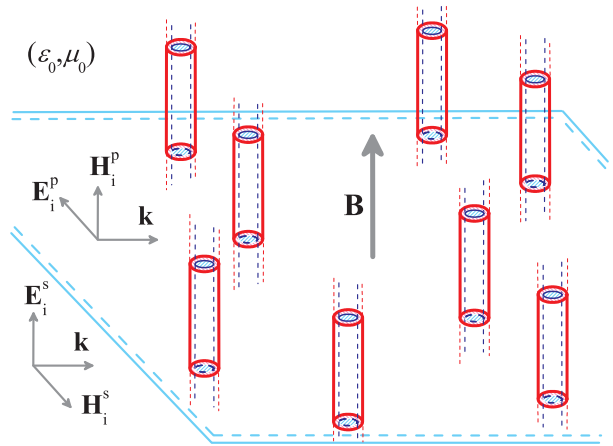}}
\caption{A two-dimensional disordered medium composed of identical parallel core-shell cylinders, embedded in the medium $(\varepsilon_0,\mu_0)$, normally irradiated with plane waves in the Voigt configuration $(\mathbf{B}\perp\mathbf{k})$.
We consider the distance among the cylinders much greater than their radius and $k\ell_{\rm sca}\gg1$ ($\ell_{\rm sca}$ being the scattering mean free path), so that the weak disorder approximation holds.
The incident polarization schemes, TM and TE modes, are indicated by $(\mathbf{E}_{\rm i}^{\rm p},\mathbf{H}_{\rm i}^{\rm p})$ and  $(\mathbf{E}_{\rm i}^{\rm s},\mathbf{H}_{\rm i}^{\rm s})$, respectively.}\label{fig2}
\end{figure}

The asymmetry parameter $\langle \cos\theta\rangle$, which is related to the transferred linear momentum in the forward direction~\cite{Bohren}, is calculated from the relationship
\begin{align}
Q_{\rm sca}\langle \cos\theta \rangle = \int_{0}^{\pi} {\rm d}\theta \frac{\partial Q(\phi)}{\partial\phi}\cos\theta\ ,
\end{align}
where $Q_{\rm sca}$ and $\partial Q(\phi)/\partial\phi$ are defined in Eqs.~(\ref{Qsca}) and (\ref{Qdif}), respectively, for a single-scattering process.

The transport mean free path is $\ell^{\star}=1/(\rho \sigma^{\star})$, where $\rho$ is the density of particles in the host medium, and $\sigma^{\star}=\sigma_{\rm ext}-\sigma_{\rm sca}\langle \cos\theta\rangle$ is the transport cross section~\cite{Ishimaru}, with $\sigma_{\rm ext}$ and $\sigma_{\rm sca}$ being the extinction and scattering cross sections, respectively.
Notice that here we take into account unavoidable losses to calculate $\ell^{\star}$, as in Refs.~\cite{Ishimaru,Ishimaru1983,Aronson}.
For lossless scatterers $\sigma_{\rm ext}=\sigma_{\rm sca}$, so that $\ell^{\star}=\ell_{\rm sca}/(1-\langle\cos\theta\rangle)$, where $\ell_{\rm sca}=1/(\rho\sigma_{\rm sca})$ is the scattering mean free path.
For a disordered 2D medium consisting of parallel cylindrical particles, as depicted in Fig.~\ref{fig2}, we obtain
\begin{align}
\frac{\ell^{\star}}{b} = \frac{\pi/2}{f_{\rm pack} \left(Q_{\rm ext} - Q_{\rm sca}\langle \cos\theta\rangle\right)}\ ,\label{ell}
\end{align}
where $f_{\rm pack}$ is the filling fraction. 
It is convenient to define the extinction mean free path: $\ell_{\rm ext}=\pi b/(2f_{\rm pack}Q_{\rm ext})$.
In this 2D case  in the Voigt configuration, the effective diffusion coefficient is $\mathcal{D}=v_{\rm E}\ell^{\star}/2$, where $v_{\rm E}$ is the energy-transport velocity.
Note that $\mathcal{D}$ does not depend explicitly on $\mathbf{B}$, which does not apply to the Faraday configuration $(\mathbf{B}||\mathbf{k})$~\cite{Maynard}. 
From the weak disorder approximation of the Bethe-Salpeter equation~\cite{Tiggelen1,Tiggelen2}, it follows that
\begin{align}
\frac{v_{\rm E}}{c} \approx \frac{1}{f_{\rm pack}(W/W_0 - 1)+1}\ , \label{ve}
\end{align}
where $c=1/\sqrt{\varepsilon_0\mu_0}$ is the velocity of light in the host medium and $W/W_0$ is the energy-enhancement factor in a single scatterer, with $W$ being the time-averaged internal EM energy~\cite{Bott,Tiago_JOSA2010,Tiago_joa,Tiago_JOSA2013}.
Equation~(\ref{ve}), originally calculated for spheres, is not restricted to low densities of scatterers~\cite{Tiggelen1} and can successfully be applied to cylinders~\cite{Ruppin_cylinder}.
In the following, we analytically calculate $W/W_0$ for a gyrotropic coated cylinder for both $p$ and $s$ waves.

\section{The exact analytic time-averaged energy within gyrotropic coated cylinders}
\label{results}

The time-averaged EM energy density within a gyroelectric and gyromagnetic medium $({\overleftrightarrow{\boldsymbol{\varepsilon}_q}},{\overleftrightarrow{\boldsymbol{\mu}_q}})$, given by Eqs.~(\ref{epsilon}) and (\ref{mu}), is
\begin{align}
\langle u_q\rangle_t &= \frac{1}{4}\bigg[\varepsilon_{q\perp}^{\rm(eff)}\left(\left|{E_{qr}}\right|^2+\left|E_{q\phi}\right|^2\right)+\varepsilon_{q||}^{\rm(eff)}\left|{E}_{qz}\right|^2\nonumber\\
&+\mu_{q\perp}^{\rm(eff)}\left(\left|{H_{qr}}\right|^2+\left|H_{q\phi}\right|^2\right)+\mu_{q||}^{\rm(eff)}\left|{H}_{qz}\right|^2\nonumber\\
 &+ 2{\rm Im}\left(\gamma_{q}^{\rm (eff)}E_{qr}E_{q\phi}^*+\eta_{q}^{\rm (eff)}H_{qr}H_{q\phi}^*\right)\bigg]\ ,\label{u}
\end{align}
where the effective energy coefficients, if the medium is weakly absorbing~\cite{Landau}, are $\varepsilon_{q\perp}^{\rm (eff)}=\partial[\omega{\rm Re}(\varepsilon_q^{\perp})]/\partial\omega$, $\varepsilon_{q||}^{\rm (eff)}=\partial[\omega{\rm Re}(\varepsilon_q^{||})]/\partial\omega$, $\gamma_{q}^{\rm (eff)}=\partial[\omega{\rm Re}(\gamma_q)]/\partial\omega$, and so forth.
Equation~(\ref{u}) is simplified whether we consider p waves ($E_{qz}=H_{q r}=H_{q\phi}=0$) or $s$ waves ($H_{qz}=E_{qr}=E_{q\phi}=0$). 

From Eq.~(\ref{u}), the corresponding time-averaged EM energy in a segment $L$ of a cylindrical shell $l_1\leq r\leq l_2$ is, therefore~\cite{Tiago_JOSA2013},
\begin{align}
    W_q=\int_{-L/2}^{L/2}{\rm d}z\int_0^{2\pi}{\rm d}\phi\int_{l_1}^{l_2}{\rm d}r\;r\langle u_q\rangle_t\ .\label{Wq}
\end{align}
If the cylindrical shell  $l_1\leq r\leq l_2$ has the same optical properties as the surrounding medium $(\varepsilon_0,\mu_0)$, it follows that
\begin{align}
    W_{0q} = \frac{\varepsilon_0}{2}|E_0|^2\pi (l_2^2-l_1^2)L\ ,
\end{align}
where $E_0$ is the electric amplitude of the incident wave.

The technical details involved in the analytical derivation of $W_q$, with $q=1$ for $(l_1,l_2)=(0,a)$ and $q=2$ for $(l_1,l_2)=(a,b)$, are given in Appendix~\ref{integrals}. 
Using the results in Appendix~\ref{integrals}, let us consider the partial contributions to the internal energy: $W_{q\perp}^{+}=\int {\rm d}^3 r \varepsilon_{q\perp}^{\rm (eff)}(|E_{r}|^2+|E_{\varphi}|^2)/4$, $W_{q||}=\int {\rm d}^3 r \varepsilon_{q||}^{\rm (eff)}|E_z|^2/4$, $W_{q\perp}^{-}= \int{\rm d}^3 r {\rm Im}(\gamma_{q}^{\rm (eff)} E_rE_{\varphi}^*)/2$, and so on.
For both $p$ or $s$ polarizations, the partial contributions to the EM energy in the cylinder have the same analytical expression, but with the corresponding Lorenz-Mie coefficients and material parameters in the equations.

From Eq.~(\ref{Wq}), we obtain for the core region ($q=1$, $l_1=0$, and $l_2=a$)
\begin{align}
    \frac{W_{1\perp}^{\pm}}{W_{01}} &= {\zeta_{1\perp}^{\pm}}\sum_{n=-\infty}^{\infty}|b_n|^2\mathcal{F}_{1,n}^{\pm (JJ)}\ ,\label{W1-perp} \\
    \frac{W_{1||}}{W_{01}} &= \zeta_{1||}\sum_{n=-\infty}^{\infty}|b_n|^2\mathcal{I}_{1,n}^{(JJ)}\ ,\label{W1-para}
		\end{align}
		where we have considered $(A,B)=(b_n,0)$ into Eqs.~(\ref{bessel1}) and (\ref{bessel2}) in Appendix~\ref{integrals} to obtain $W_{1\perp}^{+}$ and $W_{1\perp}^{-}$, respectively.
	The auxiliary functions $\mathcal{I}_{1,n}^{(JJ)}$ and $\mathcal{F}_{1,n}^{\pm(JJ)}$ are obtained from Eqs.~(\ref{integral1}) [or (\ref{integral2})] and (\ref{cal-F}), respectively, and depend on the product of the Bessel functions.
	The EM energy within the core $({\overleftrightarrow{\boldsymbol{\varepsilon}_1}},{\overleftrightarrow{\boldsymbol{\mu}_1}})$ is, therefore, 
		\begin{align}
		W_1 = W_{1\perp}^{+} + W_{1\perp}^{-} + W_{1||}\ .\label{W1}
		\end{align}
		For the cylindrical shell ($q=2$, $l_1=a$, and $l_2=b$),  we obtain
	\begin{align}
			    \frac{W_{2\perp}^{\pm}}{W_{02}} &= {\zeta_{2\perp}^{\pm}}\sum_{n=-\infty}^{\infty}\bigg\{|c_n|^2\mathcal{F}_{2,n}^{\pm (JJ)}\nonumber\\
				&+ 2{\rm Re}\left[c_nd_n^*\mathcal{F}_{2,n}^{\pm (JY)}\right]+|d_n|^2\mathcal{F}_{2,n}^{\pm (YY)} \bigg\}\ ,\label{W2-perp} \\
    \frac{W_{2||}}{W_{02}} &= \zeta_{2||}\sum_{n=-\infty}^{\infty}\bigg\{|c_n|^2\mathcal{I}_{2,n}^{(JJ)}\nonumber\\
		& + 2{\rm Re}\left[c_nd_n^*\mathcal{I}_{2,n}^{(JY)}\right]+ |d_n|^2\mathcal{I}_{2,n}^{(YY)} 	\bigg\}\ ,\label{W2-para}
\end{align}
where we have considered $(A,B)=(c_n, d_n)$ into Eqs.~(\ref{bessel1}) and (\ref{bessel2}) in Appendix~\ref{integrals} to achieve $W_{2\perp}^{+}$ and $W_{2\perp}^{-}$, respectively.
The auxiliary functions $\mathcal{F}_{2,n}^{\pm(Z\bar{Z})}$ and $\mathcal{I}_{2,n}^{(Z\bar{Z})}$ are defined in Appendix~\ref{integrals}, where $Z$ and $\bar{Z}$ are any Bessel ($J_n$) or Neumann ($Y_n$) function.
The EM energy within the shell $({\overleftrightarrow{\boldsymbol{\varepsilon}_2}},{\overleftrightarrow{\boldsymbol{\mu}_2}})$ is 
\begin{align}
W_2 = W_{2\perp}^{+} + W_{2\perp}^{-} + W_{2||}\ .\label{W2}
\end{align}

To obtain the internal energy associated with $p$ or $s$ polarization schemes one must consider  Eqs.~(\ref{tm-p}) and $(b_n^{\rm p},c_n^{\rm p}, d_n^{\rm p})$ or Eqs.~(\ref{te-s}) and $(b_n^{\rm s},c_n^{\rm s}, d_n^{\rm s})$, respectively, and  apply the relations:
\begin{align}
{\rm TM\  (p)}&\Rightarrow \begin{cases}\zeta_{q\perp}^{\rm p+}\equiv\varepsilon_{q\perp}^{\rm(eff)}/\varepsilon_0,\\
\zeta_{q\perp}^{\rm p-}\equiv \gamma_q^{\rm(eff)}/\varepsilon_0\ ,\\
\zeta_{q||}^{\rm p}\equiv |\widetilde{m}_q^{\rm p}|^2\mu_{q||}^{\rm(eff)}/\mu_0\ ;
\end{cases}\label{cases-p}\\
{\rm TE\  (s)}&\Rightarrow \begin{cases}\zeta_{q\perp}^{\rm s+}\equiv|\widetilde{m}_q^{\rm s}|^2\mu_{q\perp}^{\rm(eff)}/\mu_0,\\
\zeta_{q\perp}^{\rm s-}\equiv |\widetilde{m}_q^{\rm s}|^2\eta_q^{\rm(eff)}/\mu_0\ ,\\
\zeta_{q||}^{\rm s}\equiv \varepsilon_{q||}^{\rm(eff)}/\varepsilon_0\ .
\end{cases}\label{cases-s}
\end{align}

The energy-enhancement factor $W_{1,2}/W_0$ within the scatterer, where $W_{1,2}=W_1+W_2$ is the total internal energy and $W_0=W_{01}+W_{02}$, is
\begin{align}
    \frac{W_{1,2}}{W_0}=S^2\frac{W_{1}}{W_{01}}+\left(1-S^2\right)\frac{W_2}{W_{02}}\ ,
\end{align}
with $S=a/b$ being the aspect ratio.

In addition, since the internal field intensities are proportional to the power loss, we can write the absorption efficiency $Q_{\rm abs}^{\rm p}$ in terms of the partial energy contributions:
\begin{align}
    &\frac{{Q_{\rm abs}^{\rm p}}}{\pi y} = {\rm Im}\Bigg\{S^2\left[\frac{\varepsilon_{1}^{\perp}}{\varepsilon_{1\perp}^{\rm (eff)}}\frac{W_{1\perp}^{\rm p+}}{W_{01}} + \frac{\gamma_{1}}{\gamma_{1}^{\rm (eff)}}\frac{W_{1\perp}^{\rm p-}}{W_{01}} + \frac{\mu_1^{||}}{\mu_{1||}^{\rm (eff)}}\frac{W_{1||}^{\rm p}}{W_{01}}\right]\nonumber\\
		&+(1-S^2)\left[\frac{\varepsilon_{2}^{\perp}}{\varepsilon_{2\perp}^{\rm (eff)}}\frac{W_{2\perp}^{\rm p+}}{W_{02}} + \frac{\gamma_{2}}{\gamma_{2}^{\rm (eff)}}\frac{W_{2\perp}^{\rm p-}}{W_{02}} + \frac{\mu_2^{||}}{\mu_{2||}^{\rm (eff)}}\frac{W_{2||}^{\rm p}}{W_{02}}\right]\Bigg\}.\label{Qabs-p}
\end{align}
For $s$ waves, $Q_{\rm abs}^{\rm s}$ is obtained from Eq.~(\ref{Qabs-p}) by replacing the symbols $(\varepsilon,\gamma,\mu)$ with $(\mu,\eta,\varepsilon)$ and the label $p$ with $s$. 
It is worth mentioning that Eq.~(\ref{Qabs-p}) provides an explicit connection between the internal energy and a measurable quantity, $Q_{\rm abs}$~\cite{Bott,Tiago_JOSA2013}. 

\section{Dielectric microcylinders with magneto-optical coatings}
\label{graphics}

So far our results are general and can be applied, {e.g.}, to the study of coated gyromagnetic materials and nanowires.
Here we focus on a particular case: infinite coated gyroelectric cylinders irradiated with THz $p$ waves. 
Finite-size effects are known to weakly affect the scattering properties of cylinders provided their length is much larger than both their diameter and the incident wavelength~\cite{Finite,Bohren,Hulst}. 
Provided these conditions are met, light is mostly scattered in the plane perpendicular to the cylinder axis~\cite{Bohren}.
Some technical details regarding the calculations are provided in Appendix~\ref{Numerical}.

The  cylinder is embedded in vacuum $(\varepsilon_0,\mu_0)$ and consists of a dielectric core made of silica (SiO$_2$)  ($\varepsilon_{1} = 2.25\varepsilon_0$ and $\mu_1=\mu_0$ in the far-infrared) coated with a cylindrical shell of indium antimonide (InSb), whose dielectric tensor [Eq.~(\ref{epsilon}) for $q=2$] reads~\cite{Dai,Fan2}:
\begin{align}
\frac{\varepsilon_{2}^{\perp}(\omega,B,T)}{\varepsilon_0}&=\varepsilon_{\infty}-\frac{\omega_{\rm p}^2\left(\omega+\Gamma\imath\right)}{\omega\left[\left(\omega+\Gamma\imath\right)^2-\omega_{\rm c}^2\right]}\ ,\label{InSb}\\
\frac{\varepsilon_{2}^{||}(\omega,B,T)}{\varepsilon_0}&=\varepsilon_{\infty}-\frac{\omega_{\rm p}^2}{\omega\left(\omega+\Gamma\imath\right)}\ ,\label{InSb2}\\
\frac{\gamma_{2}(\omega,B,T)}{\varepsilon_0}&=\frac{\omega_{\rm p}^2\omega_{\rm c}}{\omega\left[\left(\omega+\Gamma\imath\right)^2-\omega_{\rm c}^2\right]}\ ,
\end{align}
where $\varepsilon_{\infty}=15.7$ is the high-frequency permittivity.
The cyclotron frequency is $\omega_{\rm c}=eB/m^{\star}$, where $e$ is the electron charge, $B$ is the external dc magnetic field, and $m^{\star}=0.015m_{\rm e}$ is the effective mass of free carriers, with $m_{\rm e}$ being the bare mass of the electron.
The plasma frequency and the collision frequency of carriers are, respectively, $\omega_{\rm p}=\sqrt{\mathcal{N}e^2/(\varepsilon_0m^{\star})}$ and $\Gamma=e/(\mu_{\rm e} m^{\star})$, where $\mathcal{N}$ is the intrinsic carrier density and $\mu_{\rm e}$ is the electron mobility.
The intrinsic carrier density (in cm$^{-3}$) in undoped InSb  is strongly dependent on the temperature and reads~\cite{Cunninghan}: 
\begin{align}
\mathcal{N}(T)\approx 5.76\times10^{14}T^{3/2}\exp\left[-0.129/\left(k_{\rm B} T\right)\right] \ ,\label{Nin} 
\end{align}
where $k_{\rm B}$ is the Boltzmann constant (in eV K$^{-1}$).
This expression, derived from the temperature variation of the Hall coefficient, agrees well with experimental data for $150~{\rm K}\leq T\leq 300~{\rm K}$~\cite{Madelung,Howells,Zimpel}; for this reason we restrict our analysis to this temperature range. 
In addition, we employ a realistic empirical expression for the electron Hall mobility (in cm$^2$V$^{-1}$s$^{-1}$)~\cite{Madelung}
\begin{align}
\mu_{\rm e}(T)\approx 7.7\times10^4\left({T}/{300}\right)^{-5/3}\ ,\label{mob}
\end{align}
which has been experimentally validated in the temperature range $150~{\rm K}\leq T\leq 300~{\rm K}$~\cite{Howells}, which we consider here. 

\begin{figure*}[htb!]
{\includegraphics[width=\columnwidth]{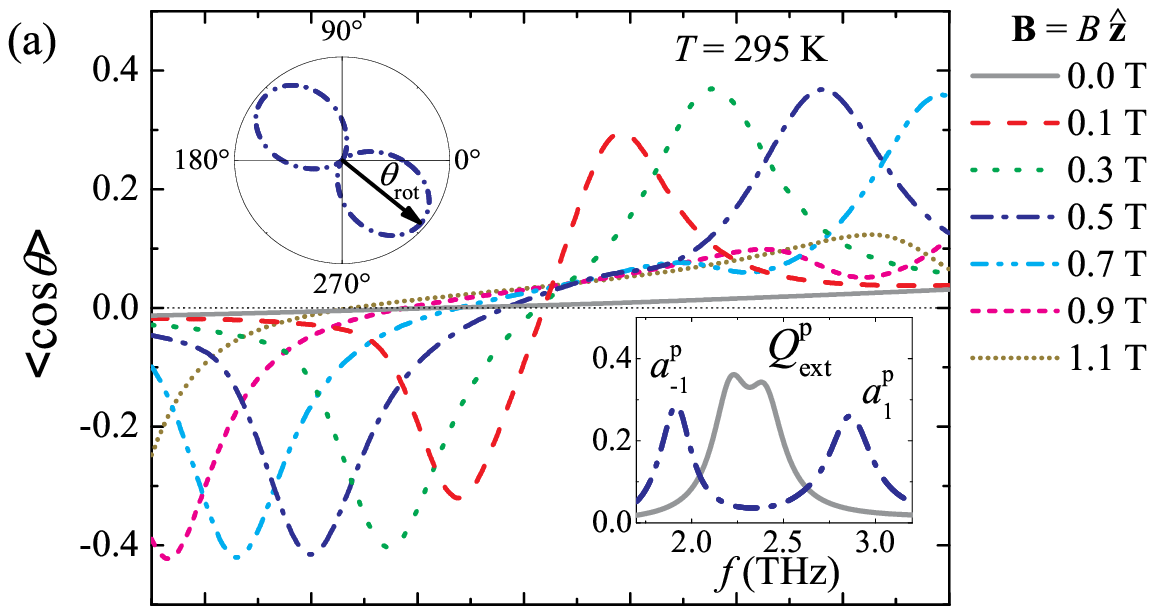}}\vspace{-0.7cm}
{\includegraphics[width=\columnwidth]{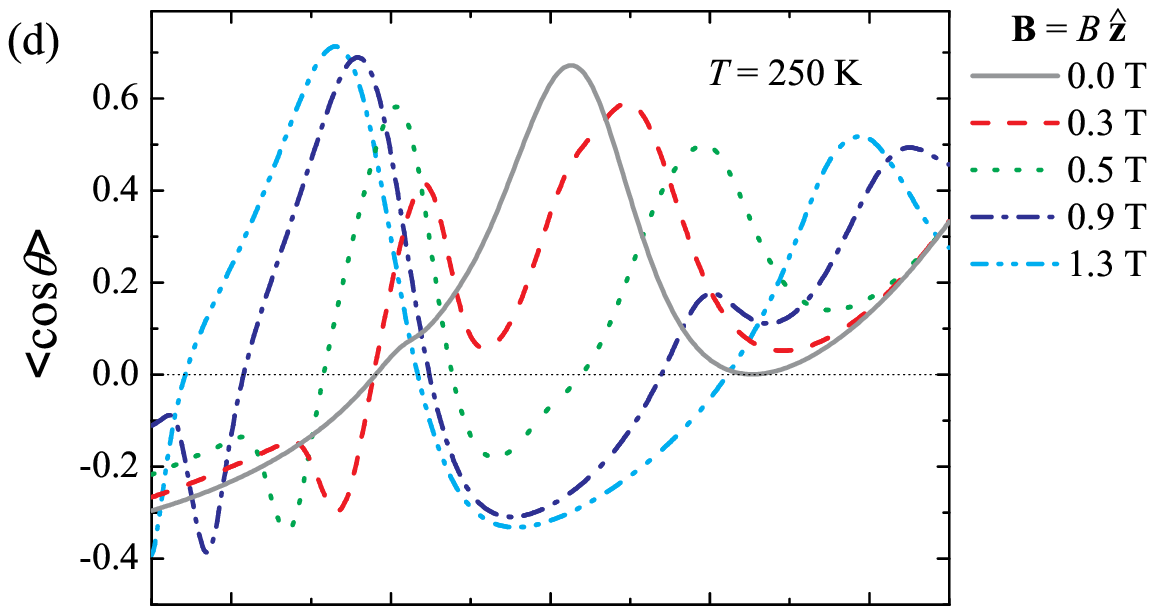}}\vspace{-0.7cm}
{\includegraphics[width=\columnwidth]{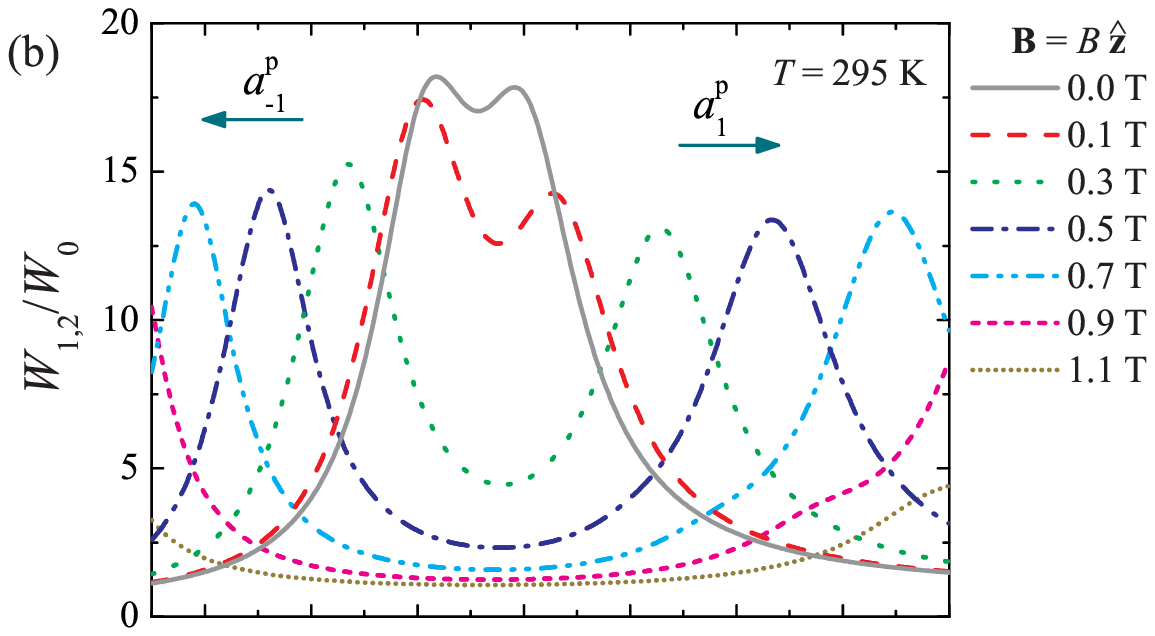}}\vspace{-0.7cm}
{\includegraphics[width=\columnwidth]{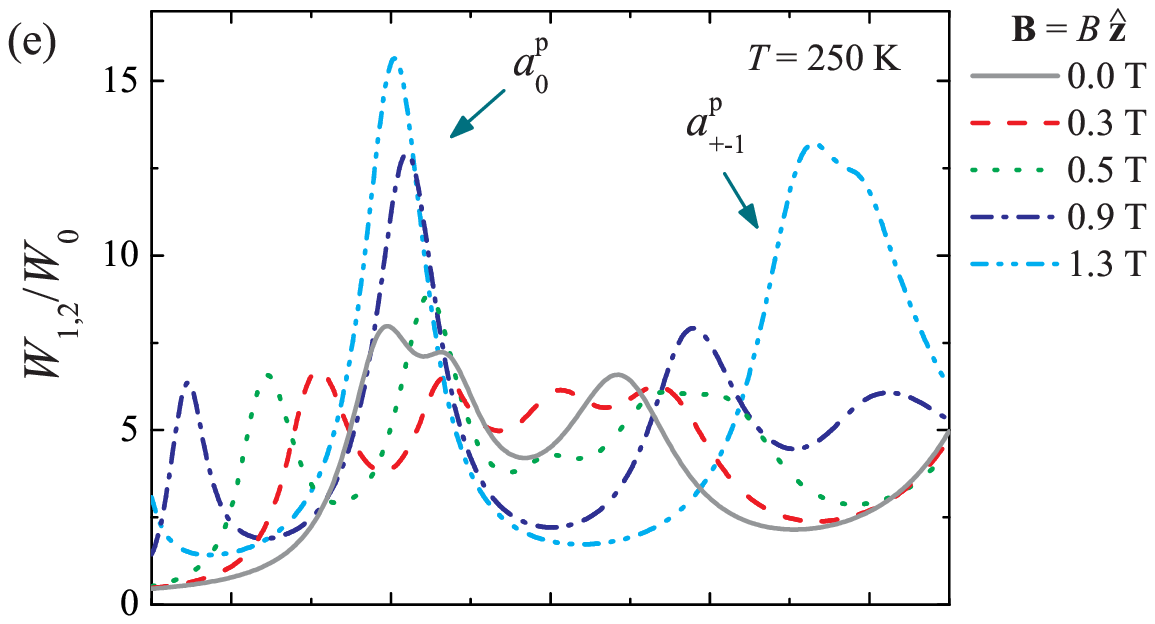}}\vspace{-0.7cm}
{\includegraphics[width=\columnwidth]{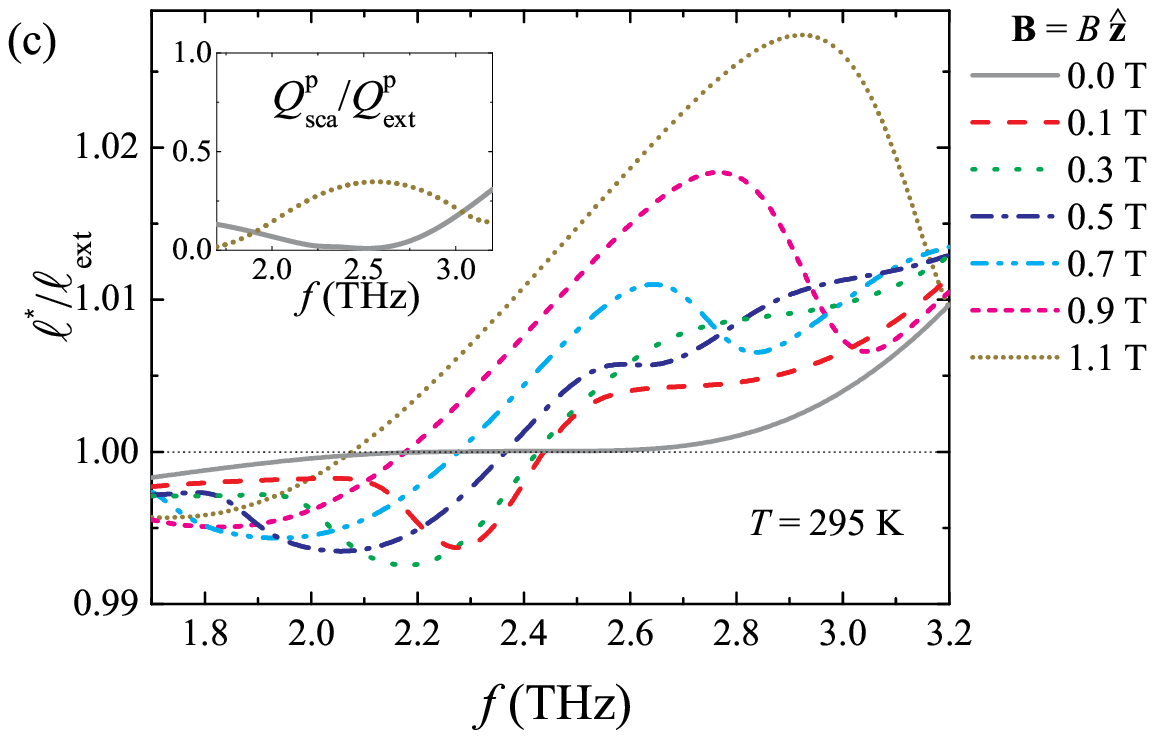}}
{\includegraphics[width=\columnwidth]{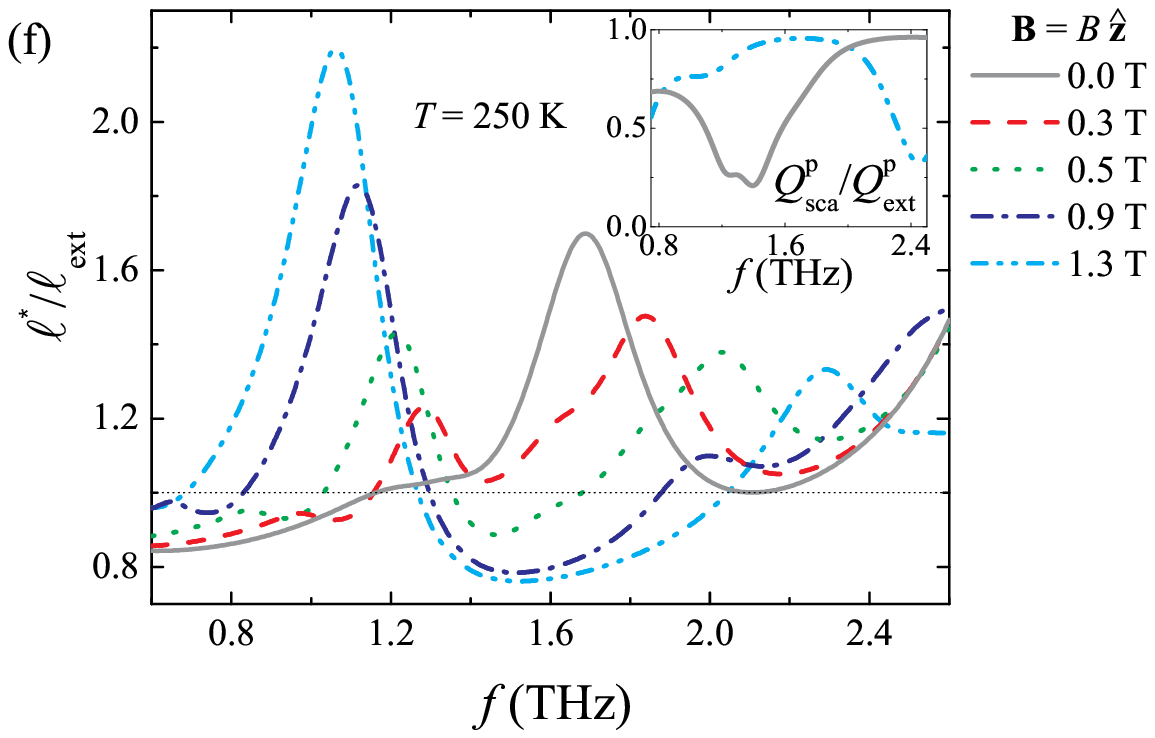}}
\caption{
Single scattering by a dielectric SiO$_2$ microcylinder $(\varepsilon_1=2.25\varepsilon_0)$ coated with undoped InSb [${\varepsilon_2}={\varepsilon_2}(\omega,B,T)$, Eqs.~(\ref{InSb})--(\ref{mob})].
The left panel (a)--(c) corresponds to room temperature ($T=295$~K), with the scatterer radius being $b=2.5$~$\mu$m ($kb\ll1$), and the aspect ratio $S=a/b=0.35$.
The right panel (d)--(f) corresponds to $T=250$~K, with the scatterer radius being $b=25$~$\mu$m ($kb\approx1$), and the aspect ratio $S=a/b=0.5$.
The system is normally irradiated with $p$ waves and is subjected to an external magnetic field $\mathbf{B}=B\hat{\mathbf{z}}$.
(a) The asymmetry parameter $\langle\cos\theta\rangle$ as a function of the frequency $f=\omega/2\pi$, for various magnetic field amplitudes $B$.
The insets show the extinction efficiency $Q_{\rm ext}$ for $B=0.0$~T ($a_1^{\rm p}=a_{-1}^{\rm p}$) and $B=0.5$~T ($a_1^{\rm p}\not=a_{-1}^{\rm p}$) and the corresponding scattering pattern for $B=0.5$~T and $f=2.0$~THz.
(b) The energy-enhancement factor $W_{1,2}/W_0$ within the core-shell cylinder.
(c) The ratio between the transport and the extinction mean free paths $\ell^{\star}/\ell_{\rm ext}$, Eq.~(\ref{ell}).
The inset shows the ratio between the scattering and the extinction efficiencies $Q_{\rm sca}^{\rm p}/Q_{\rm ext}^{\rm p}$.
For the other configuration $(kb\approx1)$, one has (d) $\langle\cos\theta\rangle$, (e) $W_{1,2}/W_0$ and (f) $\ell^{\star}/\ell_{\rm ext}$ as a function of $f$ and $B$.
}\label{fig3}
\end{figure*}

For the corresponding energy coefficients in Eq.~(\ref{u}), we consider the Loudon approach~\cite{Loudon} to deal with lossy Drude-Lorentz models~\cite{Ruppin_dispersive}:
\begin{align}
\varepsilon_{2\perp}^{\rm(eff)}(\omega) &= {\rm Re}\left[\varepsilon_{2}^{\perp}(\omega)\right] + \frac{2\omega}{\Gamma}{\rm Im}\left[\varepsilon_{2}^{\perp}(\omega)\right]\ ,\\
\gamma_{2}^{\rm(eff)}(\omega) &= {\rm Re}\left[\gamma_{2}(\omega)\right] + \frac{2\omega}{\Gamma}{\rm Im}\left[\gamma_{2}(\omega)\right]\ .
\end{align}
We recall that $\varepsilon_2^{||}(\omega)$ does not contribute to the scattering by $p$ waves.
The remaining energy coefficients are calculated by the usual Landau's formula for lossless or weakly absorbing media~\cite{Landau}. 
For non-dispersive media, it is simply the real part.

From Eqs.~(\ref{InSb}) and (\ref{InSb2}), note that $\varepsilon_2^{||}(\omega,T,B) = \varepsilon_2^{\perp}(\omega,T,0)$.
Using relations (\ref{te-s}), {i.e.}, $\varepsilon_2^{\rm s}=\varepsilon_2^{||}$ and $\mu_2^{\rm s}=\mu_0$ (with $\beta_{2}^{\rm s}=0$), one can readily verify that scattering for $s$ waves is insensitive to $\mathbf{B}$.
Indeed in the Rayleigh limit ($kb\ll1$) for $s$ waves one has $|a_0^{\rm s}|\gg|a_{\pm 1}^{\rm s}|$ for nonmagnetic scatterers~\cite{Bohren}, and hence the overall scattering response depends on the bulk resonances of the InSb associated with $\varepsilon_2^{||}$. 
For this reason, we do not consider $s$ waves in our discussion.
In addition, it is worth mentioning that oblique incidence would lead to cross-polarization coupling for higher-order modes, {\it i.e.} $s$ or $p$ waves would be scattered in a combination of both $s$ and $p$ polarization states~\cite{Finite}. 
Since the magneto-optical response is maximal for $p$ waves and vanishes for $s$ waves,  
oblique incidence would weaken the net magneto-optical effect due to radiation polarization conversion. 
For this reason, together with the fact that for normal incidence an analytical solution exists, we prefer to focus on the normal incidence case~\cite{Tiago_JOSA2013,Finite}.

In Figs.~\ref{fig3}(a)--\ref{fig3}(c), we show the asymmetry parameter $\langle \cos\theta\rangle$, the energy-enhancement factor $W_{1,2}/W_0$, and the transport mean free path $\ell^{\star}$, respectively, in a (SiO$_2$) core-shell (InSb) cylinder for $p$ waves as a function of the frequency and external magnetic field.
We set $b=2.5$~$\mu$m (with aspect ratio $S=a/b=0.35$), and room temperature ($T=295$~K).
The range of size parameters in Figs.~\ref{fig3}(a)--\ref{fig3}(c) is $0.089<kb<0.17$, so that dipole contributions  to the scattering ($n=0$ and $n=\pm1$) are dominant; in particular, the magnetic dipole contribution ($n=0$) is negligible  since $\mu_1=\mu_2=\mu_0$. 
Figure~\ref{fig3}(a) shows that the application of an external magnetic field ${\bf B}$ strongly affects the scattering directionality. 
Indeed, the presence of ${\bf B}$ breaks the scattering isotropy of dipolar scattering, in contrast to what occurs for non-Faraday-active materials in the Rayleigh regime $(kb\ll1)$. 
In these materials $\langle \cos\theta\rangle \approx 0$~\cite{Bohren} as a consequence of the typical isotropic dipolar scattering pattern, for which $Q_{\rm sca}^{\rm p}\propto|a_1^{\rm p}|^2$. 
For  magneto-optical  materials $a_{1}^{\rm p}\not=a_{-1}^{\rm p}$ for $B\not=0$ [see the inset of Fig.~\ref{fig3}(a)], leading to a strongly asymmetric, magnetic-field-dependent scattering pattern, as shown in Fig.~\ref{fig3}(a). 
In particular, the two peaks related to the dipole resonance for $B=0.0$~T are  essentially due to the presence of the dielectric core SiO$_2$.
We have verified that as $a\to 0$, only one peak remains for $a_1^{\rm p}=a_{-1}^{\rm p}$ around $f=2.4$~THz . 
Here, the dielectric core broadens the dipole resonance for $B=0.0$~T.

\begin{figure}[htb!]
{\includegraphics[width=\columnwidth]{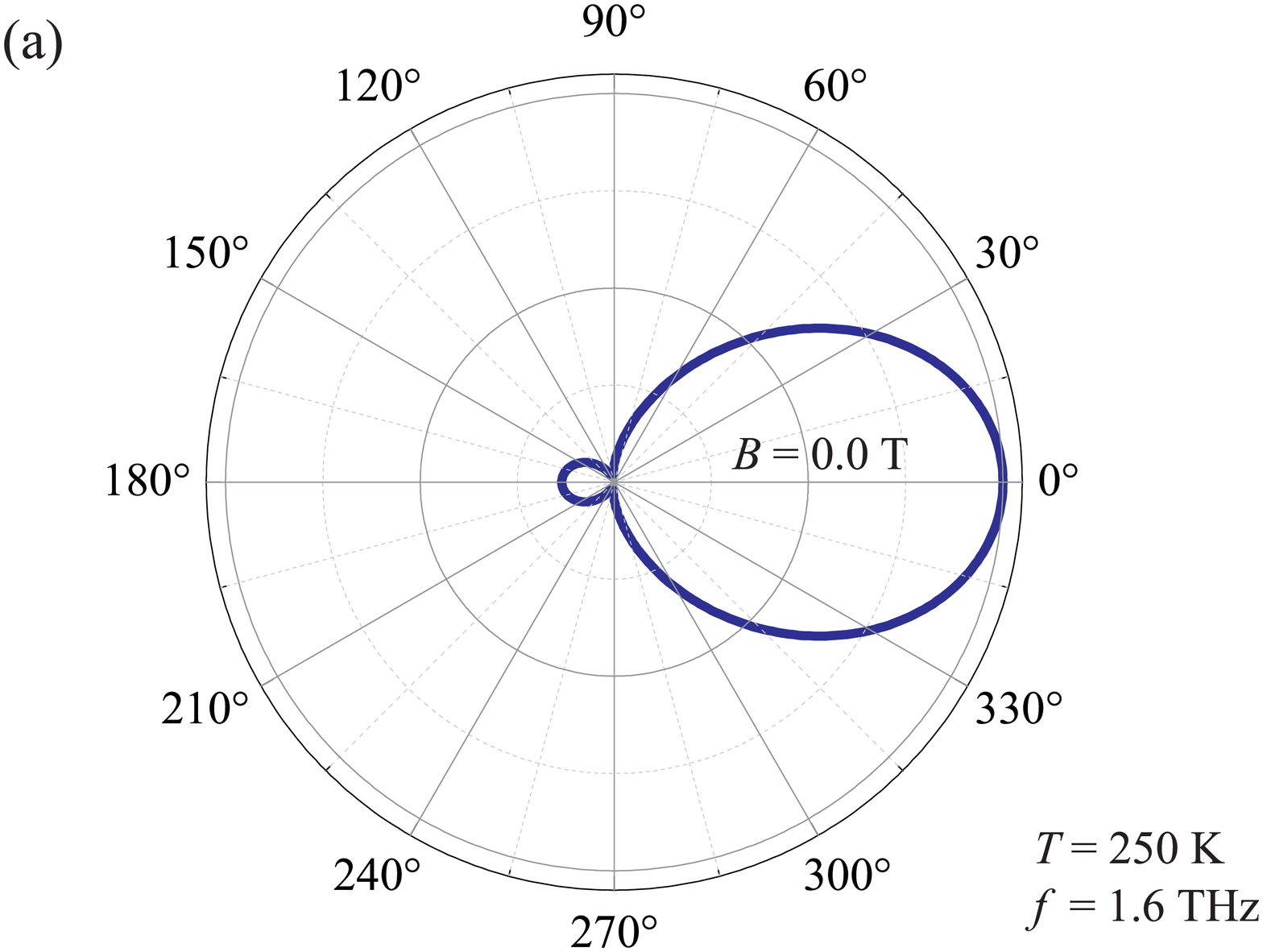}}\vspace{-0.2cm}
{\includegraphics[width=\columnwidth]{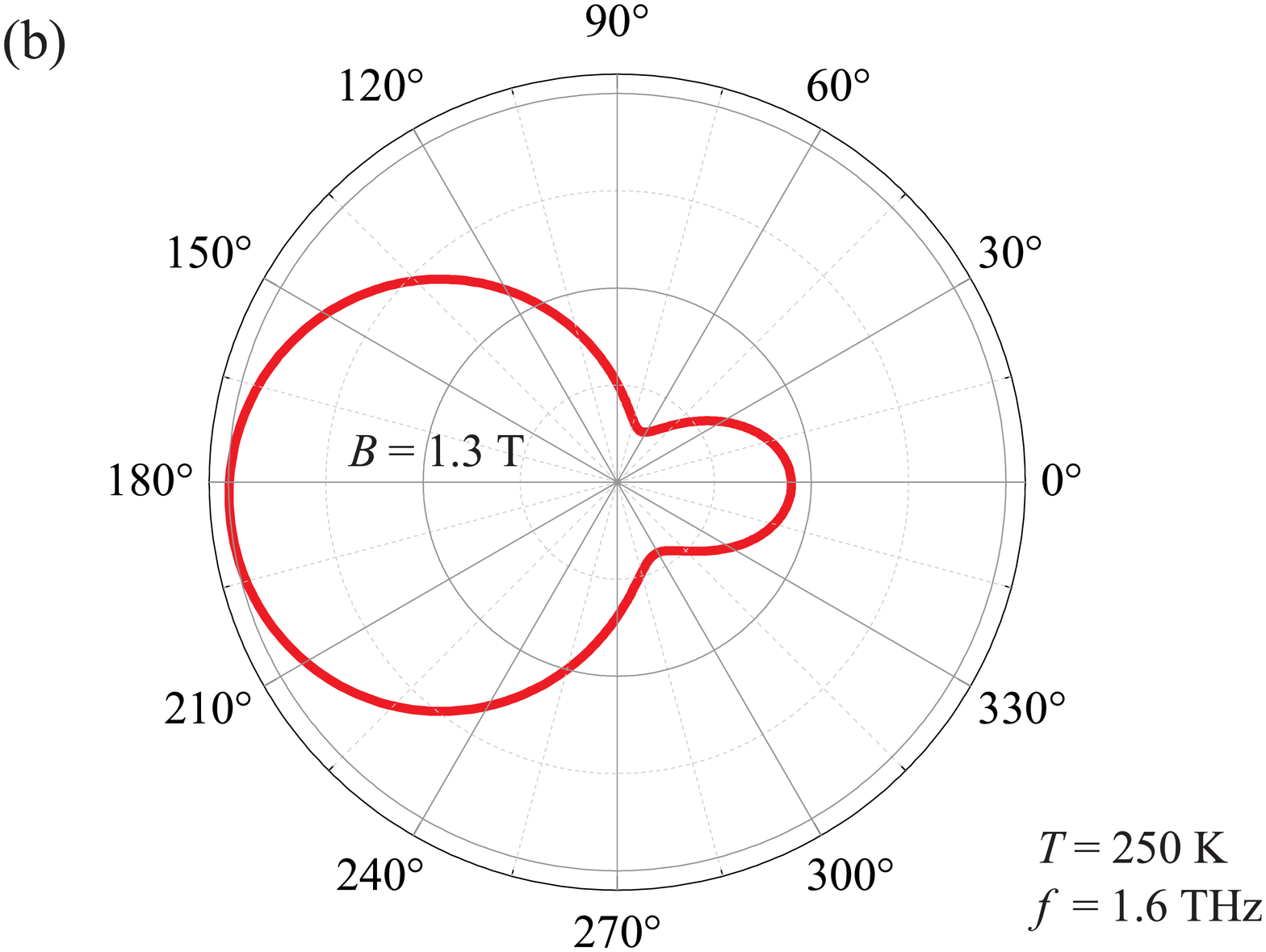}}
\caption{Normalized differential scattering efficiency $\partial Q(\phi)/\partial\phi$, Eq.~(\ref{Qdif}), for a (SiO$_2$) core-shell (InSb) cylinder  with outer radius $b=25$~$\mu$m and aspect ratio $S=a/b=0.5$, for a fixed frequency $f=1.6$~THz ($kb\approx 0.84$).
The system is normally irradiated with $p$ waves and is subjected to an external magnetic field $\mathbf{B}=B\hat{\mathbf{z}}$ and temperature $T=250$~K.
(a) $B=0.0$~T, showing preferential forward scattering ($\langle\cos\theta\rangle\approx 0.63$).
(b) $B=1.3$~T, showing  preferential backscattering ($\langle\cos\theta\rangle\approx -0.32$).
}\label{fig4}
\end{figure}

Figure~\ref{fig3}(a) reveals not only that the presence of ${\bf B}$ leads to anisotropic scattering ($\langle \cos\theta\rangle \neq 0$), but also that ${\bf B}$ induces preferential backscattering ($\langle \cos\theta\rangle < 0$), which hardly occurs in light-scattering~\cite{Bohren}. 
In particular, the conditions $\langle\cos\theta\rangle=\pm1/2$ are known as the first ($+$) and second ($-$) Kerker conditions, respectively~\cite{Kerker,Medina2012}.
Figure~\ref{fig3}(a) shows that both Kerker conditions are almost met for certain frequencies and magnetic fields due to the fact that $a_1^{\rm p}\not=a_{-1}^{\rm p}$.
The appearance of $\langle\cos\theta\rangle<0$ in the dipole approximation is explained, for $p$ waves, by the far-field interference between the coefficients $a_1^{\rm p}$ and $a_{-1}^{\rm p}$.
This breaking of the degeneracy results in a rotation of the dipolar scattering pattern [see the inset in Fig.~\ref{fig3}(a)], whose rotation angle in our notation is~\cite{Wilton_josa} 
\begin{align}
\theta_{\rm rot}\approx\frac{1}{2}\arctan\left[\frac{{\rm Im}\left(a_1^{\rm p}a_{-1}^{\rm p*}\right)}{{\rm Re}\left(a_1^{\rm p}a_{-1}^{\rm p*}\right)}\right]\ .\label{theta-rot}
\end{align}
Conversely, for the stored EM energy, we have nonvanishing interference between the electric-field components $(E_{2r}E_{2\phi}^*)$ in the $xy$ plane, as can be seen from the EM energy density expression [see Eq.~(\ref{u})].
It is worth emphasizing that, in contrast to previous  studies on directional scattering~\cite{Medina2012,Liu,Liu2013}, our approach does not rely on magnetic resonances since $a_0^{\rm p} = 0$. 
Rather, it is based on the magnetic-field dependence of electric dipolar resonances $a_1^{\rm p}$ and $a_{-1}^{\rm p}$. 

\begin{figure*}[htb!]
{\includegraphics[width=\columnwidth]{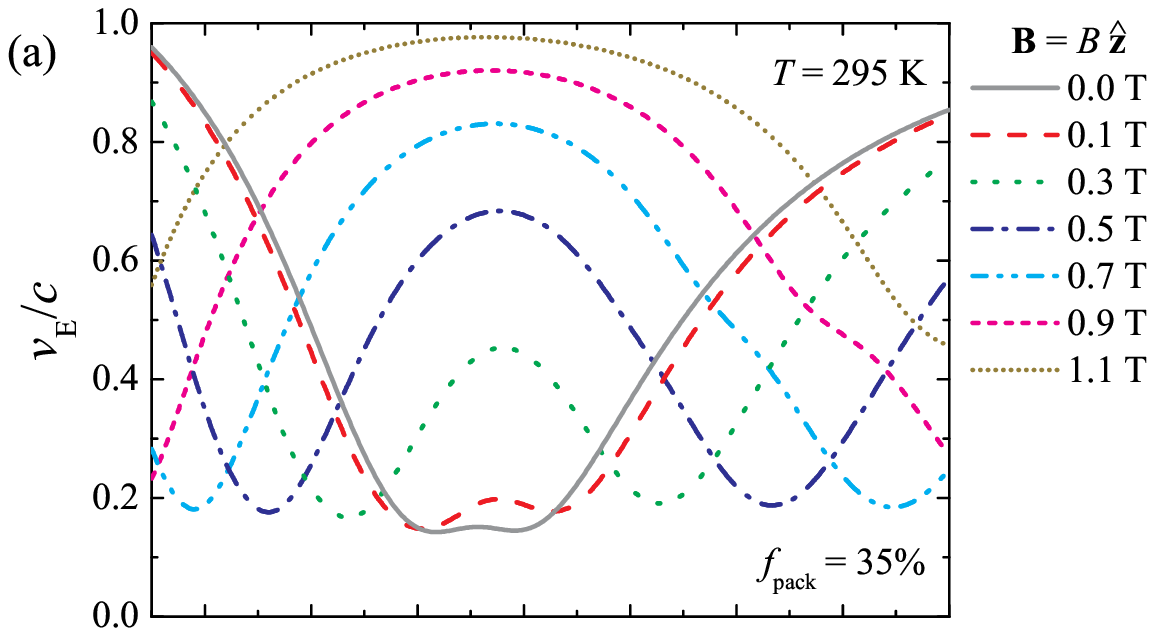}}\vspace{-0.7cm}
{\includegraphics[width=\columnwidth]{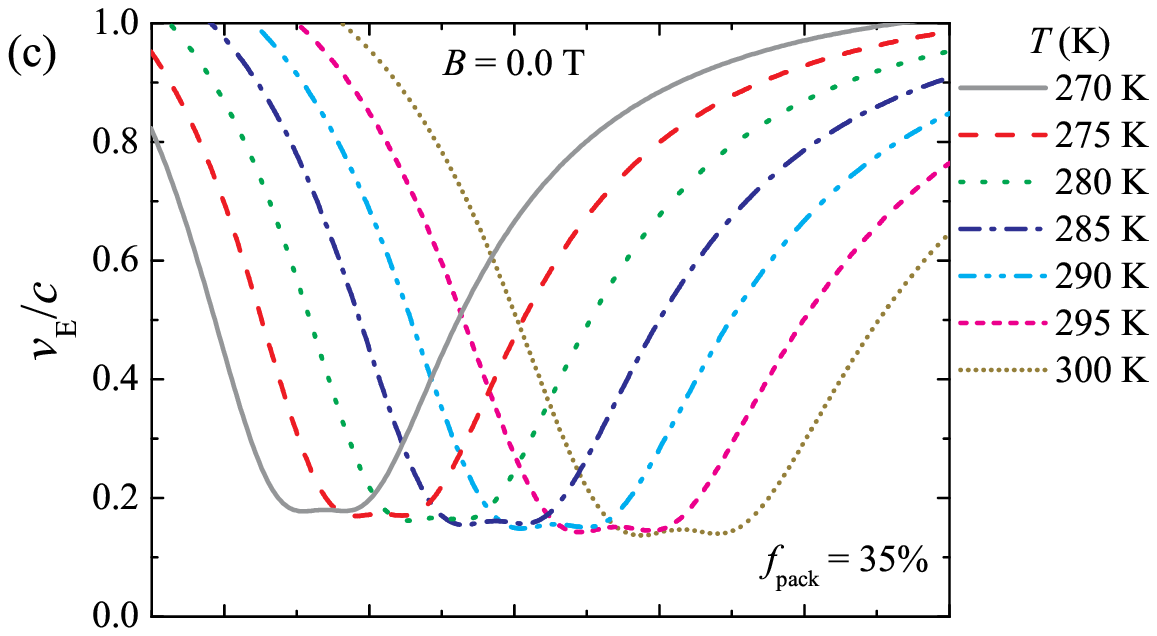}}\vspace{-0.7cm}
{\includegraphics[width=\columnwidth]{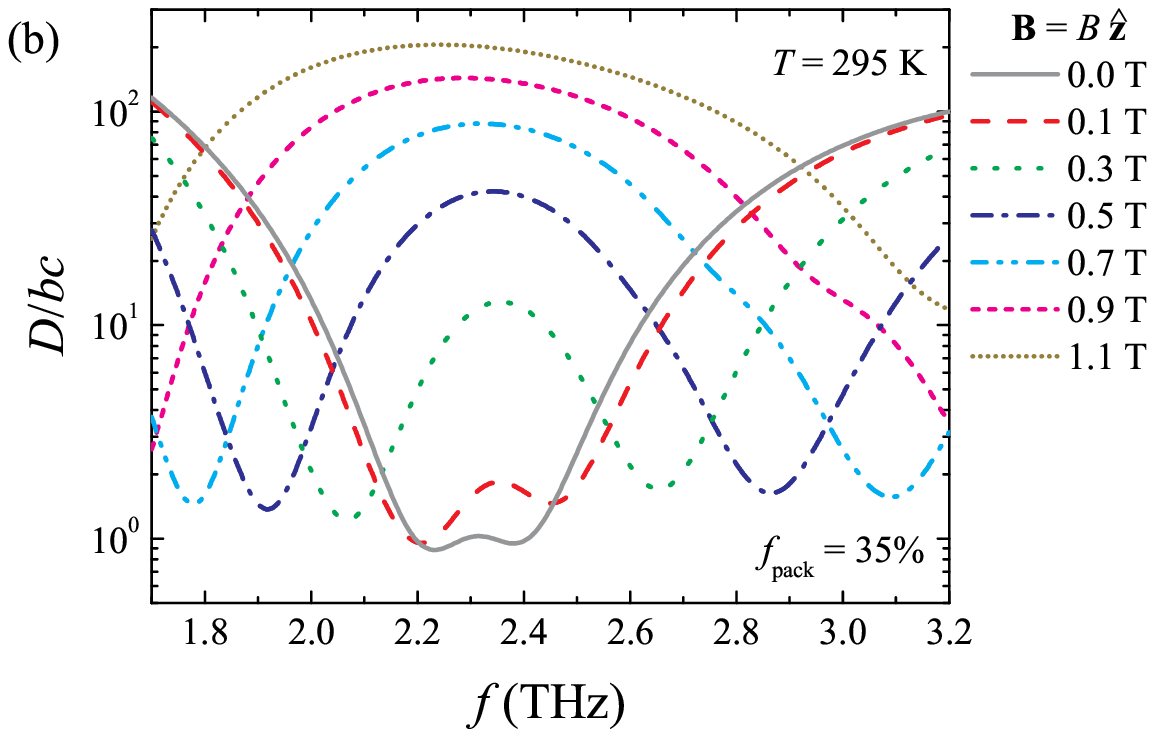}}
{\includegraphics[width=\columnwidth]{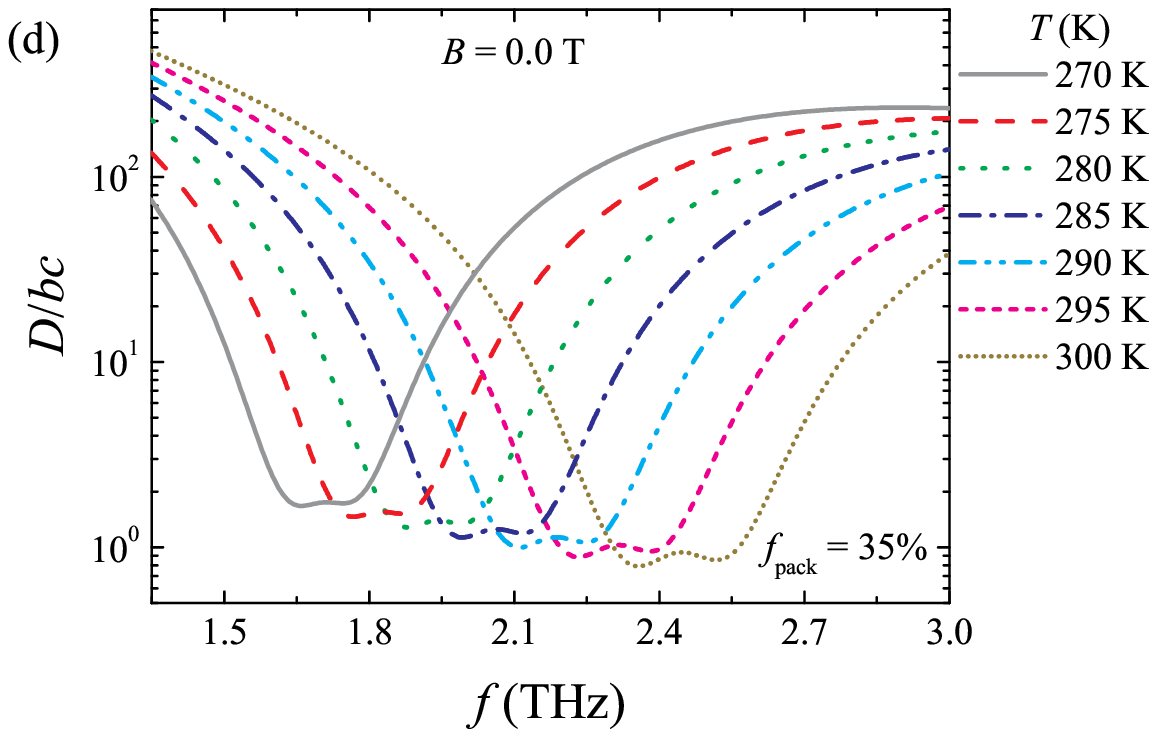}}
\caption{
Multiple scattering by parallel identical SiO$_2$ microcylinders $(\varepsilon_1=2.25\varepsilon_0)$ coated with undoped InSb $[{\varepsilon_2}={\varepsilon_2}(\omega,B,T)$, Eqs.~(\ref{InSb})--(\ref{mob})], with packing fraction $f_{\rm pack}=35\%$.
The radius of each scatterer is $b=2.5$~$\mu$m, with corresponding aspect ratio $S=a/b=0.35$.
The system is normally irradiated with $p$ waves and is subjected to an external magnetic field $\mathbf{B}=B\hat{\mathbf{z}}$ and temperature $T$.
(a) The energy-transport velocity $v_{\rm E}$ (in units of $c$), Eq.~(\ref{ve}), through the disorder medium as a function of the frequency $f=\omega/2\pi$, for various magnetic field amplitudes $B$.
(b) The corresponding diffusion coefficient $\mathcal{D}=v_{\rm E}\ell^{\star}/2$ (in units of $bc$).
(c) $v_{\rm E}$ and (d) $\mathcal{D}$ for $B=0.0$~T as a function of $f$ and temperature $T$.
}\label{fig5}
\end{figure*}

The breaking of the degeneracy in the scattering coefficients $a_{1}^{\rm p}\not=a_{-1}^{\rm p}$ in a magnetic field also shows up in the internal EM energy stored in the cylinder, $W_{1,2}$, as shown in Fig.~\ref{fig3}(b). 
In fact, by increasing $B$ the internal resonances at $a_1^{\rm p}$ and $a_{-1}^{\rm p}$ become farther apart in frequency, leading to an increasing spectral gap in  $W_{1,2}$. 
As the internal energy is proportional to the absorption cross section, $Q_{\rm abs}^{\rm p}$, for $kb\ll1$ and weak absorption~\cite{Bott,Tiago_JOSA2010}, Fig.~\ref{fig3}(b) demonstrates a novel way to externally tune EM absorption by applying an external magnetic field. 
It is worth mentioning that this effect can be achieved for moderate magnetic fields ($B \approx 0.5$~T) and that $B>0$ shifts $a_{-1}^{\rm p}$ and $a_1^{\rm p}$ to low and high frequencies, respectively; $B<0$ does the opposite. 
In Fig.~\ref{fig3}(c), the ratio $\ell^{\star}/\ell_{\rm ext}$ is shown to demonstrate that a frequency band exists below approximately 2.4~THz in which the anomalous transport regime $\ell^{\star}<\ell_{\rm ext}$ occurs. 
This band can be shifted to lower frequencies by varying $B$ and results from the negative asymmetry parameters in the same frequency range, as shown in Fig.~\ref{fig3}(a).

Figueres~\ref{fig3}(d)--\ref{fig3}(f) demonstrate that it is possible to achieve directional scattering, which can be tuned by applying an external magnetic field, beyond the Rayleigh limit. 
Indeed, in Figs.~\ref{fig3}(d)--\ref{fig3}(f) $\langle\cos\theta\rangle$, $W_{1,2}$, and $\ell^{\star}$ are calculated, respectively, for the same system but now with $b=25$~$\mu$m ($S=a/b=0.5$), and $T=250$~K. 
For the frequency range $0.6$~THz to $2.6$~THz, size parameters are $0.31<kb<1.4$, {\it i.e.} beyond the Rayleigh limit. 
In addition, by decreasing the temperature from $295$~K to $250$~K, absorption of the InSb coating also decreases significantly [see Eqs.~(\ref{Nin}) and (\ref{mob})]. 
The overall result is that for this new set of parameters absorption is small beyond the Rayleigh limit, so that $Q_{\rm sca}$ is comparable to $Q_{\rm ext}$ for $B=0.0$~T. 
In Fig.~\ref{fig3}(d), we demonstrate that $\langle\cos\theta\rangle$ becomes negative by applying ${\bf B}$ even for $kb\approx1$. 
Figure~\ref{fig3}(e) shows that the presence of ${\bf B}$ increases the magnetic dipole contribution $a_0^{\rm p}$ for low frequencies ($1.2$~THz) at the same time that it increases the electric dipole contribution $a_{\pm 1}^{\rm p}$ for high frequencies ($2.4$~THz).
The interference between electric and magnetic dipole contributions leads to a minimum in the internal energy around $f\approx1.6$~THz as ${\bf B}$ increases.
As shown in Fig.~\ref{fig3}(f), this interference induces a band ($1.3$~THz to $2.0$~THz) of anomalous scattering in which $\ell^{\star}<\ell_{\rm ext}$.
Moreover, for $B=1.3$~T, $\ell^{\star}\approx \ell_{\rm sca}/(1-\langle\cos\theta\rangle)$ since absorption becomes very small in this frequency range, as can be verified by the inset in Fig.~\ref{fig3}(f).
This implies that there exists a transport regime in which $\ell^{\star}<\ell_{\rm sca}$, with $\ell^{\star}\approx 0.8\ell_{\rm sca}$. 
It is worth mentioning that the application of the external magnetic field can suppress absorption in this frequency range, resulting in  $Q_{\rm sca}/Q_{\rm ext} \approx 1$, as shown in the inset in Fig.~\ref{fig3}(f).
It is worth emphasizing that this anomalous scattering regime, induced by the external magnetic field, occurs with the inclusion of unavoidable losses and without consideration of any positional correlation among scatterers, in contrast to Refs.~\cite{Medina2012} and \cite{Conley}, respectively.
In addition, for fixed frequency and material parameters, Fig.~\ref{fig4} shows that we can effectively tune the directional scattering pattern by applying $\mathbf{B}$. 

In Fig.~\ref{fig5}, we investigate the impact of tunable scattering anisotropy in light transport in planes composed of identical, infinitely long magneto-optical core-shell cylinders, as depicted in Fig.~\ref{fig2}. 
The parameters are the same as in Figs.~\ref{fig3}(a)--\ref{fig3}(b): $b=2.5$~$\mu$m, $S=a/b=0.35$, and $T=295$~K.
For a fixed packing fraction $f_{\rm pack}=35\%$ we calculate the energy-transport velocity $v_{\rm E}$ and the diffusion coefficient $\mathcal{D}=v_{\rm E}\ell^{\star}/2$. 
For fixed (room) temperature, Figs.~\ref{fig5}(a) and \ref{fig5}(b) show that one can effectively tune light transport with an external magnetic field. 
Indeed, Figs.~\ref{fig5}(a) and \ref{fig5}(b) reveal that the application of an external magnetic field up to $B\approx 1.0$~T leads to an increase in $v_{\rm E}$ and $\mathcal{D}$, increasing diffusion in the plane. 
In particular, as the magnetic field is increased the diffusion coefficient $\mathcal{D}$ becomes maximal at a frequency band where a minimum at $\ell^{\star}$ and $v_{\rm E}$ $(\approx 0.15c)$, and hence $\mathcal{D}$, exists for $B=0.0$~T. 
Indeed, at $B=1.1$~T, the diffusion coefficient is two orders of magnitude greater than at $B=0.0$~T. 

In Figs.~\ref{fig5}(c) and \ref{fig5}(d), we calculate $v_{\rm E}$ and $\mathcal{D}$ as a function of the frequency fixing all the aforementioned parameters, for $B=0.0$~T and for different temperatures. 
The analysis of these figures reveals that tuning the light scattering and light propagation in-plane with the temperature is also possible. 
In fact, note that by increasing the temperature from $T=270$~K to 300~K one broadens and shifts the band of minimum $v_{\rm E}$ to high frequencies, and hence the diffusion coefficient $\mathcal{D}$. 
Also, as the temperature decreases (typically for $T>220$~K), smaller magnetic fields are required to achieve a strong magneto-optical response in InSb at high frequencies, in the THz range~\cite{Fan2}.
This implies that, for $T<295$~K, smaller magnetic fields ({\it e.g.}, $B\approx 0.5$~T instead of $1.0$~T) could be applied to obtain the same energy-transport velocity enhancement exhibited in Figs.~\ref{fig5}(a) and \ref{fig5}(b).
This strong dependence on the temperature facilitates the modulation of the EM energy transport, which can be enhanced or attenuated by $B$ and shifted in frequency by varying temperature.

Although we have focused on InSb magneto-optical coatings, there are other materials that could possibly be used to achieve similar results.
As alternatives to InSb, one could use, {e.g.}, materials that are known to exhibit a low electron effective mass $m^{\star}$, and hence a high cyclotron frequency $\omega_{\rm c}$, such as InAs, HgTe, Hg$_{1-x}$Cd$_x$Te, PbTe, PbSe, PbS, and GaAs~\cite{Zawadzki,Raymond}.
In cylindrical geometry, all these materials are expected to exhibit a strong magneto-optical effect under a normal incidence of $p$ waves at high frequencies.

\section{Conclusions}
\label{conclusion}

Using the Lorenz-Mie theory, we have calculated a set of analytical expressions to completely describe the EM scattering by gyrotropic core-shell magneto-optical cylinders.
A closed analytic expression has been derived for the EM energy stored inside the cylinder. 
For concreteness, using realistic material parameters for the silica core and InSb shell, we have calculated the stored EM energy and the scattering anisotropy. 
We have shown that the application of an external magnetic field induces a drastic decrease in EM absorption in a frequency window in the THz, where absorption is maximal in the absence of the magnetic field. 
We have demonstrated not only that the scattering anisotropy can be externally tuned by applying a magnetic field, but also that it can reach negative values in the THz even in the dipolar regime. 
This is due to the fact that the external magnetic field breaks the degeneracy between the first two electric Mie scattering coefficients, which, without the magnetic field, lead to isotropic scattering. 
We have shown that this also leads to an anomalous regime of multiple light scattering in a collection of magneto-optical core-shell cylinders, in which the scattering mean free path is longer than the transport mean free path in specific ranges in the THz. 
In our approach, we have demonstrated an unprecedented degree of external control of multiple light scattering, which can be tuned by either applying an external magnetic field or varying the temperature.

\section*{Acknowledgments}

The authors thank W. J. M. Kort-Kamp for fruitful discussions at an early stage of this work and an anonymous referee for valuable comments and suggestions. The authors acknowledge the Brazilian agencies for support.
T.J.A. holds grants from FAPESP (Grant No. 2010/10052-0) and CAPES/PNPD (Grant No. 1564300), and
A.S.M. holds grants from CNPq (Grant Nos. 307948/2014-5 and 485155/2013).
F.A.P.  acknowledges The Royal Society-Newton Advanced Fellowship (Grant No. NA150208), CAPES (Grant No. BEX 1497/14-6), and CNPq (Grant No. 303286/2013-0) for financial support.

\appendix

\section{Electric and magnetic fields for $s$ waves}
\label{TEmode}

Let us briefly discuss the multipole expansions for TE mode or $s$ polarization.
According to Fig.~\ref{fig1}, we have: $[\mathbf{E}_{\rm i}^{\rm s}(r,\phi),\mathbf{H}_{\rm i}^{\rm s}(r,\phi)]=(E_0\hat{\mathbf{z}},H_0\hat{\mathbf{y}})e^{-\imath kr\cos\phi}$,  with $\mathbf{k}=-k\hat{\mathbf{x}}$.
By duality relations between electric and magnetic quantities, the EM fields for TE polarization $(\mathbf{H}_i^{\rm s}\perp\hat{\mathbf{z}})$ are readily obtained from Eqs~(\ref{Eir})-(\ref{H2z}).
First, we must redefine the material parameters according to Eq.~(\ref{te-s}), substituting $(\varepsilon_q^{\rm p},\mu_q^{\rm p},\beta_q^{\rm p})$ with $(\varepsilon_q^{\rm s},\mu_q^{\rm s},\beta_q^{\rm s})$.
The field components are then obtained by replacing $(E_{r}^{\rm p},E_{\phi}^{\rm p},H_{z}^{\rm p})$ with $(-p H_{r}^{\rm s},-p H_{\phi}^{\rm s},p^{-1} E_{z}^{\rm s})$ and $(a_n^{\rm p},b_n^{\rm p},c_n^{\rm p},d_n^{\rm p})$ with $(a_n^{\rm s},b_n^{\rm s},c_n^{\rm s},d_n^{\rm s})$, where $p = \omega\mu_0/k$ for  the incident and scattered EM fields [Eqs.~(\ref{Eir})--(\ref{Hsz})] and $p=\omega\mu_q^{\rm s}/k_q^{\rm s}$ for the internal fields [$q=1$ for Eqs.~(\ref{E1r})--(\ref{H1z}) and $q=2$ for Eqs.~(\ref{E2r})--(\ref{H2z})].
	The TE coefficients are
\begin{align}
   a_n^{\rm s} &= \frac{J_n'(y)\left[J_n(m_2^{\rm s}y) - \mathcal{A}_n^{\rm s} Y_n(m_2^{\rm s}y)\right] - \widetilde{m}_2^{\rm s}J_n(y){{\alpha}}_n^{\rm s} }
           {H_n'^{(1)}(y)\left[J_n(m_2^{\rm s}y) - \mathcal{A}_n^{\rm s} Y_n(m_2^{\rm s}y)\right]- \widetilde{m}_2^{\rm s}H_n^{(1)}(y){{\alpha}}_n^{\rm s}}\ ,\label{bn}\\
			b_n^{\rm s} &= \frac{c_n^{\rm s} \left[ J_n(m_2^{\rm s}x) - \mathcal{A}_n^{\rm s} Y_n(m_2^{\rm s}x)\right]}{ J_n(m_1^{\rm s}x)}\ ,\label{cn}\\
    c_n^{\rm s}&= \frac{2\imath/(\pi y)}{H_n'^{(1)}(y)\left[J_n(m_2^{\rm s}y) - \mathcal{A}_n^{\rm s} Y_n(m_2^{\rm s}y)\right] - \widetilde{m}_2^{\rm s}H_n^{(1)}(y){{\alpha}}_n^{\rm s}}\ ,\label{fn}\\
		d_n^{\rm s} & = -\mathcal{A}_n^{\rm s} c_n^{\rm s}\ ,\label{vn}
	\end{align}
	where the new auxiliary functions are 
	\begin{align*}
	{\alpha}_n^{\rm s} &= \mathcal{J}_n(m_2^{\rm s}y,\beta_2^{\rm s}) - \mathcal{A}_n^{\rm s} \mathcal{Y}_n(m_2^{\rm s}y,\beta_2^{\rm s})\ ,\\
		\mathcal{A}_n^{\rm s} &= \frac{\widetilde{m}_2^{\rm s} J_n(m_1^{\rm s}x)\mathcal{J}_n(m_2^{\rm s}x,\beta_2^{\rm s})-\widetilde{m}_1^{\rm s}\mathcal{J}_n(m_1^{\rm s}x,\beta_1^{\rm s})J_n(m_2^{\rm s}x)}{\widetilde{m}_2^{\rm s}J_n(m_1^{\rm s}x)\mathcal{Y}_n(m_2^{\rm s}x,\beta_2^{\rm s})-\widetilde{m}_1^{\rm s}\mathcal{J}_n(m_1^{\rm s}x,\beta_1^{\rm s})Y_n(m_2^{\rm s}x)}\ ,
		\end{align*}	
and $m_q^{\rm s}=\sqrt{\varepsilon_q^{\rm s}\mu_q^{\rm s}/(\varepsilon_0\mu_0)}$ and $\widetilde{m}_q^{\rm s}=\sqrt{\varepsilon_q^{\rm s}\mu_0/(\varepsilon_0\mu_q^{\rm s})}$.

\section{Integrals of Bessel and Neumann functions}
\label{integrals}

To calculate the stored EM energy $W_{q}$ defined in Eq.~(\ref{Wq}), we perform volume integrations involving the product of Bessel and/or Neumann functions. 
By the recurrence relations $nZ_n(\rho)=\rho Z_{n-1}(\rho)-\rho Z_n'(\rho)$ and $\rho Z_n'(\rho)=n Z_n(\rho) -\rho Z_{n+1}(\rho)$, for any cylindrical Bessel or Neumann functions $Z_n$~\cite{Watson}, we obtain
\begin{align}
    &2\left[\big|A{\mathcal{J}}_n(\rho,\beta)+B{\mathcal{Y}}_n(\rho,\beta)\big|^2+\big|A\widetilde{\mathcal{J}}_n(\rho,\beta)+B\widetilde{\mathcal{Y}}_n(\rho,\beta)\big|^2\right]\nonumber\\
    &=\left|1+\beta\right|^2\left|AJ_{n-1}(\rho)+BY_{n-1}(\rho)\right|^2\nonumber\\
    & +\left|1-\beta\right|^2\left|AJ_{n+1}(\rho)+BY_{n+1}(\rho)\right|^2\ ,\label{bessel1}\\
		&4{\rm Re}\left\{\left[A{\mathcal{J}}_n(\rho,\beta)+B{\mathcal{Y}}_n(\rho,\beta)\right]^*\left[A\widetilde{\mathcal{J}}_n(\rho,\beta)+B\widetilde{\mathcal{Y}}_n(\rho,\beta)\right]\right\}\nonumber\\
		  &=\left|1+\beta\right|^2\left|AJ_{n-1}(\rho)+BY_{n-1}(\rho)\right|^2\nonumber\\
    & -\left|1-\beta\right|^2\left|AJ_{n+1}(\rho)+BY_{n+1}(\rho)\right|^2\ .	\label{bessel2}
%    &=\left|\left(A-B\right)J_{n-1}(\rho)\right|^2+\left|\left(C-D\right)Y_{n-1}(\rho)\right|^2\\
%    &\ +2{\rm
%    Re}\left[\left(A-B\right)\left(C^*-D^*\right)J_{n-1}(\rho)Y_{n-1}(\rho^*)\right]\\
%    &\ +\left|\left(A+B\right)J_{n+1}(\rho)\right|^2+\left|\left(C+D\right)Y_{n+1}(\rho)\right|^2\\
%    &\ +2{\rm Re}\left[\left(A+B\right)\left(C^*+D^*\right)J_{n+1}(\rho)Y_{n+1}(\rho^*)\right]\;.
\end{align}
Equations~(\ref{bessel1}) and (\ref{bessel2}) are suitable for simplifying the radial integrals of the field components.
Indeed, according to Refs.~\cite{Tiago_JOSA2013}, we define, for $m_{q} \not=m_{q}^*$ ($q=\{1,2\}$), the auxiliary function
\begin{align}
    \mathcal{I}_{q ,n}^{(Z\bar{Z})}&=\frac{1}{\left(l_2^2-l_1^2\right)}\int_{l_1}^{l_2}{\rm
    d}r\ rZ_n(\rho_{q} )\bar{Z}_n(\rho_{q} ^*)\nonumber\\
    &=r^2\frac{\left[\rho_{q} ^*Z_n(\rho_{q} )\bar{Z}_n'(\rho_{q} ^*)-\rho_{q} Z_n'(\rho_{q} )\bar{Z}_n(\rho_{q} ^*)\right]}{\left(l_2^2-l_1^2\right)\left(\rho_{q} ^2-\rho_{q} ^{*2}\right)}\Bigg|_{r=l_1}^{r=l_2}\
    ,\label{integral1}
\end{align}
where $Z_n$ and $\bar{Z}_n$ are any cylindrical Bessel or Neumann functions, and $l_1,l_2\in\mathbb{R}$ are the integration limits.
Using the L'Hospital rule, if $m_{q} = m_{q} ^*$  [{\it i.e.}, ${\rm Im}(m_q)=0$], Eq.~(\ref{integral1}) can be rewritten as
\begin{align}
    {\mathcal{I}_{q ,n}^{(Z\bar{Z})}}&=\frac{r^2}{4(l_2^2-l_1^2)}\big[2Z_n(\rho_{q} )\bar{Z}_n(\rho_{q} )\nonumber\\
    &-Z_{n-1}(\rho_{q} )\bar{Z}_{n+1}(\rho_{q} )-Z_{n+1}(\rho_{q} )\bar{Z}_{n-1}(\rho_{q} )\big]\bigg|_{r=l_1}^{r=l_2}\ .\label{integral2}
\end{align}
The case $m_{q}=-m_{q}^*$ [{\it i.e.}, ${\rm Re}(m_q)=0$] is discussed in Ref.~\cite{Tiago_JOSA2013} and plays no role in our analysis.
For the sake of simplicity, we define 
\begin{align}
\mathcal{F}_{q,n}^{\pm (Z\bar{Z})} = \pm\frac{1}{2}\left[\left|1+\beta_q\right|^2\mathcal{I}_{q,n-1}^{(Z\bar{Z})} \pm \left|1-\beta_q\right|^2\mathcal{I}_{q,n+1}^{(Z\bar{Z})}\right]\ .\label{cal-F}
\end{align}

\section{Numerical calculation of the internal energy}
\label{Numerical}

Our numerical results are based on a computer code written for \textit{Scilab} 5.5.2.
For calculations, the infinite sums are truncated in $n_{\rm max}=\max(N_{\rm Mie},|m_1|y,|m_2|y)+(101+y)^{1/2}$, where $N_{\rm Mie}=y+4.05y^{1/3}+2$~\cite{Barber}.
This value guarantees the convergence of the scattering quantities \cite{Bohren}. 
In particular, it is convenient to define the internal energy for $n\geq1$ to perform numerical calculations.
To this end, we define the functions
\begin{widetext}
\begin{align*}
    \mathcal{S}_{1\perp}^{\pm} &=\frac{1}{2} \sum_{n=1}^{\infty}\bigg[\left(|b_{-n}|^2|1+\beta_1|^2\pm |b_n|^2|1-\beta_1|^2\right)\mathcal{I}_{1,n+1}^{(JJ)}+ \left(|b_{n}|^2|1+\beta_1|^2\pm |b_{-n}|^2|1-\beta_1|^2\right)\mathcal{I}_{1,n-1}^{(JJ)}\bigg]\ ,\\
		\mathcal{S}_{1||}&=\sum_{n=1}^{\infty}\left(|b_{-n}|^2+|b_n|^2\right)\mathcal{I}_{1,n}^{(JJ)}\ ,\\		
		    \mathcal{S}_{2\perp}^{\pm} &=\frac{1}{2} \sum_{n=1}^{\infty}\bigg\{\left(|c_{-n}|^2|1+\beta_2|^2\pm |c_n|^2|1-\beta_2|^2\right)\mathcal{I}_{2,n+1}^{(JJ)}+ \left(|c_{n}|^2|1+\beta_2|^2\pm |c_{-n}|^2|1-\beta_2|^2\right)\mathcal{I}_{2,n-1}^{(JJ)}\nonumber\\
		&+\left(|d_{-n}|^2|1+\beta_2|^2\pm |d_n|^2|1-\beta_2|^2\right)\mathcal{I}_{2,n+1}^{(YY)}+ \left(|d_{n}|^2|1+\beta_2|^2\pm |d_{-n}|^2|1-\beta_2|^2\right)\mathcal{I}_{2,n-1}^{(YY)}\nonumber\\
		&+2{\rm Re}\bigg[\left(c_{-n}d_{-n}^*|1+\beta_2|^2\pm c_nd_n^*|1-\beta_2|^2\right)\mathcal{I}_{2,n+1}^{(JY)}+ \left(c_nd_{n}^*|1+\beta_2|^2\pm c_{-n}d_{-n}^*|1-\beta_2|^2\right)\mathcal{I}_{2,n-1}^{(JY)}\bigg]\bigg\}\ ,\\
		\mathcal{S}_{2||}&=\sum_{n=1}^{\infty}\bigg\{\left(|c_{-n}|^2+|c_n|^2\right)\mathcal{I}_{2,n}^{(JJ)}+\left(|d_{-n}|^2+|d_n|^2\right)\mathcal{I}_{2,n}^{(YY)}+2{\rm Re}\left[\left(c_{-n}d_{-n}^*+c_nd_n^*\right)\mathcal{I}_{2,n}^{(JY)}\right]\bigg\}\ .
\end{align*}
\end{widetext}
With this set of expressions, Eqs.~(\ref{W1-perp})--(\ref{W2}) can be rewritten for both $p$ and $s$ waves, reading
\begin{widetext}
\begin{align}
    \frac{W_{1\perp}^{+}}{W_{01}} &= \zeta_{1\perp}^+\left[|b_0|^2\left(1+|\beta_1|^2\right)\mathcal{I}_{1,1}^{(JJ)} + \mathcal{S}_{1\perp}^{+}\right]\ ,\label{W1-p}\\
    \frac{W_{1\perp}^{-}}{W_{01}} &= -\zeta_{1\perp}^{-}\left[2|b_0|^2{\rm Re}(\beta_1)\mathcal{I}_{1,1}^{(JJ)} + \mathcal{S}_{1\perp}^{-}\right]\ ,\\
    \frac{W_{1||}}{W_{01}} &= \zeta_{1||}\bigg[|b_0|^2\mathcal{I}_{1,0}^{(JJ)}+\mathcal{S}_{1||}\bigg]\ ,\\
    \frac{W_{2\perp}^{+}}{W_{02}} &= \zeta_{2\perp}^+\bigg\{|c_0|^2\left(1+|\beta_2|^2\right)\mathcal{I}_{2,1}^{(JJ)}+ 2{\rm Re}\left[c_0d_0^*\left(1+|\beta_2|^2\right)\mathcal{I}_{2,1}^{(JY)}\right]+|d_0|^2\left(1+|\beta_2|^2\right)\mathcal{I}_{2,1}^{(YY)} +\mathcal{S}_{2\perp}^{+}\bigg\}\ ,\\
    \frac{W_{2\perp}^{-}}{W_{02}} &= -\zeta_{2\perp}^{-}\bigg\{
		2|c_0|^2{\rm Re}(\beta_2)\mathcal{I}_{2,1}^{(JJ)} +4{\rm Re}\left[c_0d_0^{*}{\rm Re}(\beta_2)\mathcal{I}_{2,1}^{(JY)}\right]+2|d_0|^2{\rm Re}(\beta_2)\mathcal{I}_{2,1}^{(YY)}+ \mathcal{S}_{2\perp}^{-}\bigg\}\ ,\\
    \frac{W_{2||}}{W_{02}} &= \zeta_{2||}\bigg\{
		|c_0|^2\mathcal{I}_{2,0}^{(JJ)}+2{\rm Re}\left[c_0d_0^{*}\mathcal{I}_{2,0}^{(JY)}\right]+|d_0|^2\mathcal{I}_{2,0}^{(YY)}+\mathcal{S}_{2||}\bigg\}\ .\label{W2-p}
\end{align}
\end{widetext}


\begin{thebibliography}{99}

\bibitem{Bohren}
C. F. Bohren and D. R. Huffman,
{\it Absorption and Scattering of Light by Small Particles}
(Wiley, New York, 1983).

\bibitem{Huschka} 
R. Huschka, J. Zuloaga, M. W. Knight, L. V. Brown, P. Nordlander, and N. J. Halas, 
%``Light-induced release of DNA from gold nanoparticles: nanoshells and nanorods,''
J. Am. Chem. Soc. {\bf 133}, 12247 (2011).

\bibitem{Alu2008}
A. Al\`u and N. Engheta,
%``Multifrequency optical invisibility cloak with layered plasmonic shells,''
{Phys. Rev. Lett.} {\bf 100}, 113901 (2008).

\bibitem{Edwards}
B. Edwards, A. Al\`u, M. G. Silveirinha, and N. Engheta,
%``Experimental verification of plasmonic cloaking at microwave frequencies with metamaterials,''
Phys. Rev. Lett. {\bf 103}, 153901 (2009).
 
\bibitem{Wilton-prl} 
W. J. M. Kort-Kamp, F. S. S. Rosa, F. A. Pinheiro, and C. Farina, 
%``Tuning plasmonic cloaks with an external magnetic field,''  
{Phys. Rev. Lett.} {\bf 111}, 215504 (2013).

\bibitem{Tribelsky}
M. I. Tribelsky, A. E. Miroshnichenko, and Y. S. Kivshar,
%``Unconventional Fano resonances in light scattering by small particles,''
Europhys. Lett. {\bf 97}, 44005 (2012).

%\bibitem{Tiago_pra1}
%T. J. Arruda, A. S. Martinez, and F. A. Pinheiro,
%``Unconventional Fano effect and off-resonance field enhancement in plasmonic coated spheres,''
%{Phys. Rev. A} {\bf 87}, 043841 (2013).
%
%\bibitem{Tiago_pra2}
%T. J. Arruda, A. S. Martinez, and F. A. Pinheiro, 
%``Tunable multiple Fano resonances in magnetic single-layered core-shell particles,''
%Phys. Rev. A {\bf 92}, 023835 (2015).

\bibitem{Tiago_pra}
T. J. Arruda, A. S. Martinez, and F. A. Pinheiro,
%``Unconventional Fano effect and off-resonance field enhancement in plasmonic coated spheres,''
{Phys. Rev. A} {\bf 87}, 043841 (2013); {\bf 92}, 023835 (2015)

\bibitem{Chen_gao}
H. L. Chen and L. Gao,
%``Tunablity of the unconventional Fano resonances in coated nanowires with radial anisotropy,''
Opt. Express {\bf 21}, 23619 (2013).

\bibitem{Jordi}
J. Sancho-Parramon and D. Jelovina,
%``Boosting Fano resonances in single layered concentric core-shell particles,''
Nanoscale {\bf 6}, 13555 (2014).

\bibitem{Miromagnetic} 
A. E. Miroshnichenko, B. Luk'yanchuk, S. A. Maier, and Y. S. Kivshar 
%``Optically induced interaction of magnetic moments in hybrid metamaterials,''
ACS Nano  {\bf 6}, 837 (2012).

\bibitem{Kuznetsov} 
A. I. Kuznetsov, A. E. Miroshnichenko, Y. H. Fu, J. Zhang, and B. Luk'yanchuk,
%``Magnetic light,'' 
Sci. Rep. {\bf 2}, 492 (2012).

\bibitem{Ruan} 
Z. Ruan and S. Fan, 
 %``Superscattering of light from subwavelength nanostructures,''
Phys. Rev. Lett., {\bf 105}, 013901 (2010).  
 
\bibitem{Liu} 
W. Liu, A. E. Miroshnichenko, D. N. Neshev, and Y. S. Kivshar,
 %``Broadband unidirectional scattering by magneto-electric core-shell nanoparticles,''
ACS Nano {\bf 6}, 5489 (2012).

\bibitem{Liu2013} 
W. Liu, A. E. Miroshnichenko, R. F. Oulton, D. N. Neshev, O. Hess, and Y. S. Kivshar, 
%``Scattering of core-shell nanowires with the interference of electric and magnetic resonances,''
 Opt. Lett. {\bf 38}, 2621 (2013).

\bibitem{Boris} 
B. S. Luk'yanchuk and V. Ternovsky, 
%``Light scattering by thin wire with surface plasmon resonance: bifurcations of the Poynting vector field,''
Phys. Rev. B {\bf 73}, 235432 (2006).

\bibitem{Etxarri} 
A. Garcia-Etxarri, R. Gomez-Medina, L. S. Froufe-Perez, C. Lopez, L. Chantada, F. Scheffold, J. Aizpurua, M. Nieto-Vesperinas, and J. J. Saenz,
  %``Strong magnetic response of submicron Silicon particles in the infrared,''
 Opt. Express  {\bf 19}, 4815 (2011).

\bibitem{Kerker} 
M. Kerker,  D. S. Wang, and L. Giles 
%``Electromagnetic scattering by magnetic spheres,"  
J. Opt. Soc. Am. {\bf 73}, 765 (1983).

\bibitem{Staude} 
I. Staude, A. E. Miroshnichenko, M. Decker, N. T. Fofang, S. Liu, E. Gonzales, J. Dominguez, T. S. Luk, D. N. Neshev, I. Brener, and Y. Kivshar,
%``Tailoring directional scattering through magnetic and electric resonances in subwavelength silicon nanodisks,''
 ACS Nano {\bf 7}, 7824 (2013).

\bibitem{Vesperinas} 
M. Nieto-Vesperinas, R. Gomez-Medina, and J. J. Saenz, 
J. Opt. Soc. Am. A {\bf  18}, 54 (2011).

\bibitem{Coenen} 
T. Coenen, F. Bernal Arango, A. Femius Koenderink, and A. Polman, 
Nat. Commun. {\bf 5}, 3250 (2014).

\bibitem{Fu} 
Y. H. Fu, A. I. Kuznetsov, A. E. Miroshnichenko, Y. F. Yu, and B. Luk'yanchuk, 
Nat. Commun. {\bf 4}, 1527 (2013).

\bibitem{Zambrana} 
X. Zambrana-Puyalto, I. Fernandez-Corbaton, M. L. Juan, X. Vidal, and G. Molina-Terriza, 
Opt. Lett. {\bf  38}, 1857 (2013).

\bibitem{Hancu} 
I. M. Hancu, A. G. Curto, M. Castro-Lopez, M. Kuttge, and N. F. Van Hulst, 
Nano Lett. \textbf{14}, 166 (2014).

\bibitem{Person} 
S. Person, M. Jain, Z. Lapin, J. J. Saenz, G. Wicks, and L. Novotny
 %``Demonstration of zero optical backscattering from single nanoparticles,"
Nano Letters {\bf 13}, 1806 (2013).

\bibitem{Geffrin} 
J. M. Geffrin,	B. Garcia-Camara,	R. Gomez-Medina,	P. Albella,	L.S. Froufe-Perez,	C. Eyraud, A. Litman,	R. Vaillon,	F. Gonzalez,	M. Nieto-Vesperinas,	J. J. Saenz, and F. Moreno
%``Magnetic and electric coherence in forward- and back-scattered electromagnetic waves by a single dielectric subwavelength sphere,''
Nat. Comm.  {\bf  3}, 1171 (2012). 

\bibitem{Alaee} 
R. Alaee, R. Filter, D. Lehr, F. Lederer, and C. Rockstuhl, 
%``A generalized Kerker condition for highly directive nanoantennas,''
Opt. Lett. {\bf  40}, 2645 (2015).

\bibitem{Li} 
Y. Li, M. Wan, W. Wu, Z. Chen, P. Zhan, and Z. Wang,
%``Broadband zero-backward and near-zero-forward scattering by metallo-dielectric core-shell nanoparticles.'' 
Sci. Rep. {\bf  5} 12491 (2015).

%\bibitem{prl1}
%F. A. Pinheiro, A. S. Martinez, and L. C. Sampaio,
%``New effects in light scattering in disordered media and coherent backscattering cone: system of magnetic particles,''
%Phys. Rev. Lett. {\bf 84}, 1435-1438 (2000).
%
%\bibitem{prl2}
%F. A. Pinheiro, A. S. Martinez, and L. C. Sampaio,
%``Vanishing of energy transport velocity and diffusion constant of electromagnetic waves in disordered magnetic media,''
%Phys. Rev. Lett. {\bf 85}, 5563-5566 (2000).

\bibitem{Wilton_josa} 
W. J. M. Kort-Kamp, F. S. S. Rosa, F. A. Pinheiro, and C. Farina, 
%``Molding the flow of light with a magnetic field: plasmonic cloaking and directional scattering,'' 
{J. Opt. Soc. Am. A} {\bf 31}, 1969 (2014); W. J. M. Kort-Kamp, arXiv:1505.02333. 

\bibitem{Hall}
D. Lacoste, B. A. van Tiggelen, G. L. J. A. Rikken, and A. Sparenberg,
%Optics of a Faraday-active Mie sphere
J. Opt. Soc. Am. A {\bf 15}, 1636 (1998).

\bibitem{Felipe_prl}
F. A. Pinheiro, A. S. Martinez, and L. C. Sampaio,
%`New effects in light scattering in disordered media and coherent backscattering cone: system of magnetic particles,''
Phys. Rev. Lett. {\bf 84}, 1435 (2000); {\bf 85}, 5563 (2000).

\bibitem{Medina2012}
R. Gomez-Medina, L. S. Froufe-Perez, M. Yepez, F. Scheffold, M. Nieto-Vesperinas, and J. J. Saenz,
%``Negative scattering asymmetry parameter for dipolar particles: Unusual reduction of the transport mean free path and radiation pressure,''
Phys. Rev. A {\bf  85}, 035802 (2012).

\bibitem{Conley} 
G. M. Conley, M. Burresi, F. Pratesi, K. Vynck, and D. S. Wiersma,
Phys. Rev. Lett. {\bf 112}, 143901 (2014).

\bibitem{Madelung}
O. Madelung,
{\it Physics of III-V Compounds},
(Wiley, New York, 1964).
% Chap. 4, p. 115

\bibitem{Zimpel}
M. Oszwalldowski and M. Zimpel,
%``Temperature dependence of intrinsic carrier concentration and density of states effective mass of heavy holes in InSb,''
J. Phys. Chem. Solids {\bf 49}, 1179 (1988).

\bibitem{Howells}
S. C. Howells and L. A. Schlie,
%``Transient terahertz reflection spectroscopy of undoped InSb from 0.1 to 1.1 THz,''
Appl. Phys. Lett. {\bf 69}, 550 (1996).

\bibitem{Monzon}
J. C. Monzon and N. J. Damaskos, 
%``Two-dimensional scattering by a homogeneous anisotropic rod,''
IEEE Trans. Antennas Propag. {\bf AP-34}, 1243 (1986).

\bibitem{Tiago_JOSA2010}
T. J. Arruda and A. S. Martinez,
%``Electromagnetic energy within a magnetic sphere,''
J. Opt. Soc. Am. A {\bf 27}, 992 (2010); {\bf 27}, 1679 (2010).

%\bibitem{tiago-coated-cylinder}
%T. J. Arruda, A. S. Martinez, and F. A. Pinheiro,
%``Electromagnetic energy within coated cylinders at oblique incidence and applications to graphene coatings,''
%{J. Opt. Soc. Am. A} {\bf 31}, 1811-1819 (2013).
%
%\bibitem{tiago-omni}
%T. J. Arruda, A. S. Martinez, and F. A. Pinheiro, 
%``Omnidirectional absorption and off-resonance field enhancement in dielectric cylinders coated with graphene layers,''
%J. Opt. Soc. A {\bf 32}, 943-948 (2015).

%\bibitem{tiago-active}
%T. J. Arruda, F. A. Pinheiro, and A. S. Martinez,
%``Electromagnetic energy within single-resonance chiral metamaterial spheres,''
%J. Opt. Soc. Am. A. {\bf 30}, 1205-1212 (2013).

\bibitem{Tiago_joa}
T. J. Arruda, F. A. Pinheiro, and A. S. Martinez,
%``Electromagnetic energy within coated spheres containing dispersive metamaterials,''
J. Opt. {\bf 14}, 065101 (2012).

\bibitem{Tiago_JOSA2013}
T. J. Arruda, A. S. Martinez, and F. A. Pinheiro,
%``Electromagnetic energy within coated cylinders at oblique incidence and applications to graphene coatings,''
{J. Opt. Soc. Am. A} {\bf 31}, 1811 (2013);  {\bf 30}, 1205 (2013); {\bf 32}, 943 (2015).

\bibitem{Kivshar_cross}
B. S. Luk'yanchuk, A. E. Miroshnichenko, and Y. S. Kivshar,
J. Opt. {\bf 15}, 073001 (2013).

\bibitem{Hulst}
H. C. van de Hulst,
{\it Light Scattering by Small Particles}
(Dover, New York, 1981).

\bibitem{Tiggelen1}
M. P. van Albada, B. A. van Tiggelen, A. Lagendijk, and A. Tip,
%``Speed of propagation of classical waves in strongly scattering media,''
Phys. Rev. Lett. {\bf 66}, 3132 (1991).

\bibitem{Tiggelen2}
B. A. van Tiggelen, A. Lagendijk, M. P. van Albada, and A. Tip,
%``Speed of light in random media,''
Phys. Rev. B {\bf 45}, 12233 (1992).

\bibitem{Ishimaru}
A. Ishimaru,
{\it Wave Propagation and Scattering in Random Media}
(Academic Press, New York, 1978).

\bibitem{Ishimaru1983}
A. Ishimaru, Y. Kuga, R. L.-T. Cheung, and K. Shimizu,
%``Scattering and diffusion of a beam wave in randomly distributed scatterers,''
J.  Op. Soc. Am. {\bf 73}, 131 (1983).

\bibitem{Aronson}
R. Aronson and N. Corngold,
%``Photon diffusion coefficient in an absorbing medium,''
J. Opt. Soc. Am. A {\bf 16}, 1066 (1999).

\bibitem{Maynard}
A. S. Martinez and R. Maynard,
%``Faraday effect and multiple scattering of light,''
Phys. Rev. B {\bf 50}, 3714 (1994).

\bibitem{Bott}
A. Bott and W. Zdunkowski,
%``Electromagnetic energy within dielectric spheres,''
J. Opt. Soc. Am. A {\bf 4}, 1361 (1987).

\bibitem{Ruppin_cylinder}
R. Ruppin,
%``Electromagnetic energy inside an irradiated cylinder,''
J. Opt. Soc. Am. A {\bf 15}, 1891 (1998).

\bibitem{Landau}
L. D. Landau and E. M. Lifshits,
{\it Electrodynamics of Continuous Media}
(Pergamon Press, Oxford, 1984).

\bibitem{Finite}
A. Al\`u, D. Rainwater, and A. Kerkhoff,
%``Plasmonic cloaking of cylinders: finite length, oblique illumination and cross-polarization coupling,''
New J. Phys. {\bf 12}, 103028 (2010).


%------------------------------------------------------------------
%InSb articles

\bibitem{Dai}
X. Dai, Y. Xiang, S. Wen, and H. He,
%``Thermally tunable and omnidirectional terahertz photonic bandgap in the one-dimensional photonic crystals containing semiconductor InSb,''
J. Appl. Phys. {\bf 109}, 053104 (2011).

\bibitem{Fan2}
S. Chen, F. Fan, X. He, M. Chen, and S. Chang,
%``Multifunctional magneto-metasurface for terahertz one-way transmission and magnetic field sensing,''
App. Opt.  {\bf 54}, 9177 (2015).

\bibitem{Cunninghan}
R. W. Cunningham and J. B. Gruber, 
%``Intrinsic concentration and heavy-hole mass in InSb,''
J. Appl. Phys. {\bf 41}, 1804 (1970).

\bibitem{Loudon}
R. Loudon,
%``The propagation of electromagnetic energy through an absorbing dielectric,''
J. Phys. A: Gen. Phys. {\bf 3}, 233 (1970).

\bibitem{Ruppin_dispersive}
R. Ruppin,
%``Electromagnetic energy density in a dispersive and absorptive material,''
Phys. Lett. A. {\bf 299}, 309 (2002).

\bibitem{Zawadzki}
W. Zawadzki, 
Adv. Phys. {\bf 23}, 435 (1974).

\bibitem{Raymond}
A. Raymond, J. L. Robert, and C. Bernard, 
J. Phys. C: Solid State Phys. {\bf 12}, 2289 (1979).

\bibitem{Watson}
G. N. Watson,
{\it A Treatise on the Theory of Bessel Functions}
(Cambridge Univ. Press, Cambridge, 1958).

\bibitem{Barber}
P. W. Barber and S. C. Hill,
{\it Light Scattering by Particles: Computational Methods} 
(World Scientific, Singapore, 1990).

\end{thebibliography}
\end{document}